\documentclass[10pt,journal,compsoc]{IEEEtran}

\ifCLASSOPTIONcompsoc
\usepackage[nocompress]{cite}
\else
\usepackage{cite}
\fi

\usepackage{booktabs} 
\usepackage{xspace}
\usepackage{xcolor}
\usepackage{comment}
\usepackage{multirow}
\usepackage{amsmath}
\usepackage{amsfonts}
\usepackage{makecell}
\usepackage{subfigure}
\usepackage{footnote}
\usepackage{graphicx}
\usepackage{hyperref}

\makesavenoteenv{tabular}
\makesavenoteenv{table}

\newcommand{\etal}{et al.\@\xspace}
\newcommand{\tabincell}[2]{\begin{tabular}{@{}#1@{}}#2\end{tabular}}

\newcommand{\setblstr}[2][.]{%
	\renewcommand{\baselinestretch}{#2}%
	\ifx#1.%
	\else%
	\renewcommand{\arraystretch}{#1}%
	\fi%
}
\setblstr[1.0]{.985}

\newcounter{tightlistcnt}
\newenvironment{tightitemize}{%
	\begin{list}{\textbullet}{\usecounter{tightlistcnt}%
			\topsep 0in
			\partopsep 0in
			\itemsep .2em
			\parsep 0in
			\leftmargin 0em
			\rightmargin 0in
			\listparindent 0em
			\itemindent .75em
			\labelsep .75em
			\labelwidth 0in
		}%
	}{%
	\end{list}%
}

\newenvironment{tightenumerate}{%
	\begin{list}{\arabic{tightlistcnt}.}{\usecounter{tightlistcnt}
			\topsep 0in
			\partopsep 0in
			\itemsep .2em
			\parsep 0in
			\leftmargin 0.0em
			\rightmargin 0in
			\listparindent 0em
			\itemindent .75em
			\labelsep .75em
			\labelwidth 0in
		}%
	}{%
	\end{list}%
}

\begin{document}
	
\title{Modeling Dynamic User Preference via Dictionary Learning for Sequential Recommendation}

\author{
	Chao~Chen,~\IEEEmembership{Member,~IEEE,}
	Dongsheng~Li,~\IEEEmembership{Member,~IEEE,}
	Junchi~Yan,~\IEEEmembership{Member,~IEEE,}
	Xiaokang Yang,~\IEEEmembership{Fellow,~IEEE}

\IEEEcompsocitemizethanks{
	\IEEEcompsocthanksitem
	Chao Chen is with School of Electronic Information and Electrical Engineering, and MoE Key Lab of Artificial Intelligence, AI Institute, Shanghai Jiao Tong University, Shanghai, 200240, P. R. China.
	E-mail: chao.chen@sjtu.edu.cn.
	\IEEEcompsocthanksitem
	Dongsheng Li (correspondence author) is a senior researcher with Microsoft Research Asia, Shanghai, and an adjunct professor with 
	School of Computer Science, Fudan University, Shanghai, 201203, P. R. China.
	E-mail: dongshengli@fudan.edu.cn.
	\IEEEcompsocthanksitem
	Junchi Yan (correspondence author) is with Department of Computer Science and Engineering, and MoE Key Lab of Artificial Intelligence, AI Institute, Shanghai Jiao Tong University, Shanghai, 200240, P. R. China.
	E-mail: yanjunchi@sjtu.edu.cn.
	\IEEEcompsocthanksitem
	Xiaokang Yang is with MoE Key Lab of Artificial Intelligence, AI Institute, Shanghai Jiao Tong University, Shanghai, 200240, P. R. China.
	E-mail: xkyang@sjtu.edu.cn.
	\IEEEcompsocthanksitem Dongsheng Li and Junchi Yan are the corresponding authors.
}
}

\IEEEtitleabstractindextext{
\begin{abstract}
Capturing the dynamics in user preference is crucial to better predict user future behaviors because user preferences often drift over time. Many existing recommendation algorithms -- including both shallow and deep ones -- often model such dynamics independently, i.e., user static and dynamic preferences are not modeled under the same latent space, which makes it difficult to fuse them for recommendation. This paper considers the problem of embedding a user's sequential behavior into the latent space of user preferences, namely \textit{translating sequence to preference}. To this end, we formulate the sequential recommendation task as a dictionary learning problem, which learns: 1) a shared \textit{dictionary matrix}, each row of which represents a partial signal of user dynamic preferences shared across users; and 2) a \textit{posterior distribution estimator} using a deep autoregressive model integrated with Gated Recurrent Unit (GRU), which can select related rows of the dictionary to represent a user's dynamic preferences conditioned on his/her past behaviors. Qualitative studies on the Netflix dataset demonstrate that the proposed method can capture the user preference drifts over time and quantitative studies on multiple real-world datasets demonstrate that the proposed method can achieve higher accuracy compared with state-of-the-art factorization and neural sequential recommendation methods. The code is available at \url{https://github.com/cchao0116/S2PNM-TKDE2021}.
\end{abstract}

\begin{IEEEkeywords}
	Collaborative filtering, sequential recommendation, dynamic preference, dictionary learning
\end{IEEEkeywords}
}

\maketitle

\IEEEdisplaynontitleabstractindextext

\IEEEpeerreviewmaketitle
	
\IEEEraisesectionheading{\section{Introduction}\label{sec:intro}}
\IEEEPARstart{I}{n} recommender systems, user preferences often drift over time due to various reasons, e.g., changes in life~\cite{Hossein15}, experience growth~\cite{McAuley13}, etc. However, many existing collaborative filtering (CF) algorithms~\cite{mnih2007probabilistic,koren2008factorization} summarize a user's historical records into a single latent vector, which may lose the dynamic preference drifts and lead to suboptimal recommendation accuracy~\cite{McAuley13,Hossein15,chen2018sequential}. To remedy this, many techniques have been adopted to model user dynamic preferences in collaborative filtering. For instance, Rendle~\etal~\cite{RendleWWW10} and He~\etal~\cite{HeICDM16} both adopted Markov chains to model local patterns among items. Zheng~\etal\cite{zheng2016neural} used the neural autoregressive distribution estimator (NADE) to model the conditional probability of the next item given a user's historical ratings. Recently, several works~\cite{HidasiICLR16,rrn2017,baral2018close} used recurrent neural networks (RNN) to embed previously purchased products for current interest prediction. Due to the recent advances of deep learning, these RNN-based methods have achieved promising results in sequential recommendation. Our case study shows that RNN-based methods~\cite{rrn2017,zheng2016neural} that consider sequential patterns exhibit better accuracy than state-of-the-art matrix approximation methods~\cite{koren2008factorization,chen2017gloma,mrma} without considering sequential information on the Netflix prize dataset.

However, existing RNN-based methods~\cite{rrn2017,zheng2016neural,YuSIGIR16}, which exhibit the advantage by capturing temporal or sequential user behaviors, may not be suitable for modeling and capturing the long-term impact of previously purchased items on the future one~\cite{chen2018sequential}. To remedy this, many works~\cite{wang2018neural,chen2018sequential,huang2018isrkemn,ren2019lifelong,ying2018sequential,kang2018self,zhang2019feature} proposed to use external memory networks (EMN)~\cite{weston2014memory,sukhbaatar2015end} and attention mechanism~\cite{chorowski2015attention,vaswani2017attention}, where shorter path between any positions in the sequence makes the long-term impact easier to learn. Despite their effectiveness, the ordinal information of historical items is usually not explicitly considered in these works, which may lead to suboptimal performance because the sequential patterns contained in the sequence of user behaviors may be neglected. In addition, the sequential patterns may be conceptually different from user preferences, i.e., in different feature spaces, and thereby it may increase the complexity of learning the downstream estimators when user preferences are directly combined with the intermediate outputs of recurrent/memory/attention networks.


In this paper, we propose a dictionary learning-based approach to model both long-term static user preferences and short-term dynamic user preferences under the same latent space. More specifically, we learn a dictionary from scratch and each row of the dictionary can be regarded as a basis  representing a partial signal of user dynamic preferences shared across all users. Then, the dynamic preference of each user can be modeled by a linear combination of all the rows in the dictionary. To achieve adaptive linear combination on different users / sequences, we propose a deep autoregressive model integrated with Gated Recurrent Unit (GRU) to learn features from user sequential behaviors, which can generate the weights for the linear combination to form the dynamic preference of each user. After obtaining the dynamic preference of each user, we use an additive mechanism to fuse the dynamic and static preferences for predicting the future interests of each user.

The main contributions of this work are summarized as follows:
\begin{enumerate}
	\item To the best of our knowledge, this is the first work that tackles
	the sequential recommendation problem via dictionary learning,
	which can model user static preferences and dynamic preferences under
	the same latent space to achieve simpler fusion, i.e., a simple additive mechanism
	can achieve decent performance.
	\item Sequence-to-preference neural machine (S2PNM) is proposed to
	translate the sequential behaviors of users into dynamic user preferences using
	a deep autoregressive model integrated with GRU.
	\item Empirical studies on multiple real-world datasets demonstrate that
	S2PNM can significantly outperform state-of-the-art factorization and neural
	sequential recommendation methods in recommendation accuracy.
\end{enumerate}

\section{Problem Formulation}
\label{sec:case_study}
This section first formulates the targeted problem and then presents a case study to show the challenge faced by existing CF methods without considering sequential information.
	
\subsection{Matrix Factorization (MF)}
Matrix factorization-based collaborative filtering algorithms have recently achieved superior performance in both rating prediction task~\cite{koren2008factorization,chen2017gloma} and top-N recommendation task~\cite{he2016fast,Hu08}. Given a user-item rating matrix $R^{m\times{n}}$ with $m$ users and $n$ items, we denote $r_{ij|t}$ as the score rated by user $i$ on item $j$ at time $t$. Then for each user, there always exists an item sequence $\{x_1, \dots, x_T\}$ ($T$ will vary for different users). Traditional MF methods, e.g., regularized SVD (RSVD)~\cite{paterek2007improving}, are popular due to simple implementation and superior accuracy compared to other kinds of methods. More formally, these methods generally minimize the sum-squared error between the rating matrix $R$ and its low-rank recovery $\hat{R}=UV^{\top}$ with $\ell_{2}$ regularization as follows:
\begin{align}\min_{U,V} || \mathbb{I} \circ (R - UV^{\top})|| + \lambda_{1}||U|| + \lambda_{2}||V|| \label{eqn:prob_svd},\end{align}
where $U\in{\mathbb{R}^{m\times{k}}}, V\in{\mathbb{R}^{n\times{k}}}$ are the latent factors of users and items respectively, and $\mathbb{I}_{ij}$ is an indicator function that equals to $1$ when $R_{ij}$ is observed and equals to $0$ otherwise. $||U|| := \sqrt{\sum_{ij} U_{ij}^2}$ is the Frobenius norm of $U$. 

\subsection{Case Studies}
Traditional MF methods~\cite{koren2008factorization,chen2017gloma,mrma}
did not consider the sequential information, and thus may yield suboptimal
performance in real scenarios in which user future interests
are predicted on their historical behaviors.
Here, we conduct a case study on the Netflix dataset to demonstrate that
state-of-the-art MF methods indeed underperform in sequential-based
data splitting protocol. More specifically, we adopt two different
data splitting protocols on the Netflix dataset to study how the prediction
accuracy varies: 1) random splitting, which is the most widely adopted protocol
in classical collaborative filtering literature~\cite{koren2008factorization,rrn2017};
and 2) sequential-based splitting, in which we split the dataset based on
the chronological order to simulate real recommendation scenarios.
To make a fair comparison, we keep the ratio of training and test sets
as $9$:$1$ for both data splitting protocols.

\begin{table}[tb!]
	\caption{Recommendation RMSE of SVD++~\cite{koren2008factorization},
		GLOMA~\cite{chen2017gloma}, MRMA~\cite{mrma}, RRN~\cite{rrn2017},
		NADE~\cite{zheng2016neural} and the proposed S2PNM method, as well as
		the relative performance gains over the baseline -- SVD++ on the Netflix
		dataset with two different data splitting protocols (i.e., \textit{split by random}
		and \textit{split by time}). Note that we set the rank as 300 for SVD++,
		GLOMA and MRMA, 300 hidden units for RRN and S2PNM, and 1000 hidden units for NADE.}
	\centering
	{
		\begin{tabular}{ l | c | c }\hline
			Method&  {spit-by-random}    &  {spit-by-time}        \\\hline\hline
			SVD++   &    0.80701  $(+ 0.00\%)$  &     0.89267 $(+ 0.00\%)$     \\
			GLOMA   &    0.80132  $(+ 0.70\%)$  &     0.89326 $(- 0.06\%)$     \\
			MRMA    &    0.79940  $(+ 0.94\%)$  &    0.89210 $(+ 0.06\%)$     \\\hline
			RRN	  &	   0.80443 $(+ 0.32\%)$ & 0.89014 $(+ 0.28\%)$       \\
			NADE      &    0.80243 $(+ 0.56\%)$  &     0.88876 $(+ 0.43\%)$     \\
			{\bf S2PNM} &  {\bf 0.78481  $(+ 2.75\%)$}    &	{\bf 0.87301  $(+ 2.20\%)$}     \\\hline
	\end{tabular}}
	\label{tbl:cmp_casestudy}
\end{table}

Table~\ref{tbl:cmp_casestudy} summaries the recommendation accuracy
in terms of RMSE on one baseline method (SVD++~\cite{koren2008factorization}),
two state-of-the-art MF methods (MRMA~\cite{mrma} and GLOMA~\cite{chen2017gloma}),
two sequential recommendation methods (RRN~\cite{rrn2017} and NADE~\cite{zheng2016neural}),
and meanwhile the proposed S2PNM method. As shown in the results,
non-sequential methods (MRMA and GLOMA) achieved higher accuracy
than sequential methods (RRN and NADE) on random splitting protocol
but achieved lower accuracy on sequential-based splitting protocol.
This confirms that capturing the correlations between user historical
behavior and his/her future interests can help to achieve better recommendations
in real scenarios. In addition, when comparing the numbers between the
second column and the third column, we can see that the accuracies of
GLOMA and MRMA are on par with the baseline method -- SVD++ in the
sequential-based splitting protocol whereas they significantly outperform
SVD++ in the random splitting protocol. This indicates that conclusions
made under the unrealistic random splitting protocol may not hold
in the realistic setting, and therefore it is necessary to design and
evaluate recommendation algorithms in the harder but more realistic
sequential-based splitting protocol.

\section{The Proposed Sequence-to-Preference Neural Machine}
\label{sec:m_arch}
\begin{figure*}[tbh!]
	\centering
	\includegraphics[width=.85\textwidth]{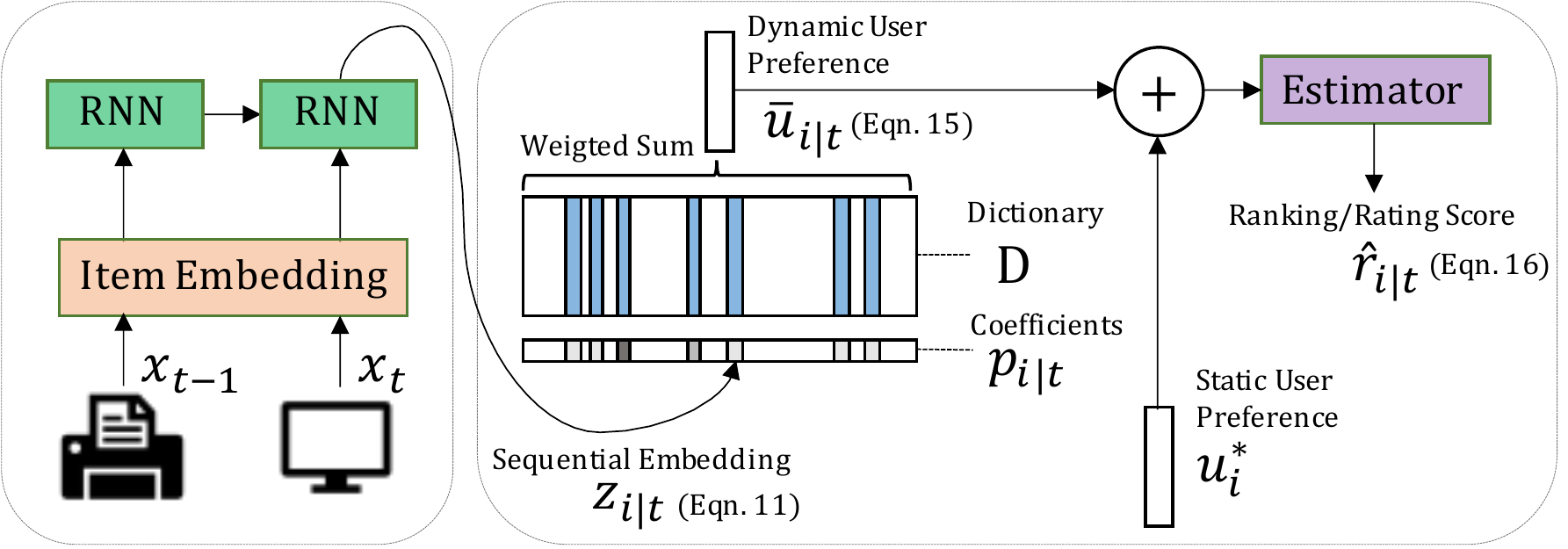}
	\caption{Architecture of sequence to preference (Seq2Pref) neural network. We first feed the $i^\mathrm{th}$ user's historical items $\{x_1, \dots, x_T\}$ to RNNs to embed sequential dependencies into a hidden vector $z_{i|t}$, then the dynamic user preference $\bar{u}_{i|t}$ is translated from $z_{i|t}$ by weighting the rows of dictionary $D$ differently, assigning higher (lower) weights $p_{i|t}$ to rows that contribute more (less) to match users' current interests. Finally, an estimator can provide the rating/ranking score $\hat{r}_{ij|t}$ for recommendations with dynamic user preference $\bar{u}_{i|t}$, static user preference $u_i^\ast$, item embedding $v_j$, overall mean rating $b_g$, and biases of $i^\mathrm{th}$ user and $j^\mathrm{th}$ item $b_i$ and $b_j$ .\label{fig:m_shallow}}
\end{figure*}

This section first presents the main building block of S2PNM -- the Seq2Pref network in detail. Then, we discuss the choice of prediction function for recommendation scores. After that, we discuss the optimization problems for two recommendation tasks. At last, we discuss the parallel training of S2PNM.

\subsection{Seq2Pref Network}
\label{sec:s2pblock}

\subsubsection{Dictionary Learning}
Dictionary learning aims to learn a set of vectors capable of succinct expression of the targeted events, i.e., a linear combination of the rows of the dictionary learned from the data can succinctly represent any piece of the input data~\cite{lee2001algorithms,mairal2009online}. In this paper, we try to learn a dictionary that can construct sufficient representations of user dynamic preferences and meanwhile embed user dynamic preferences into the same latent space of user static preferences.

More formally, we propose to learn a feature dictionary $D \in \mathbb{R}^{d_{\mathrm{dict}} \times d_\mathrm{user}}$ such that the dynamic user preference of user $i$ at time $t$ --- $\overline{u}_{i|t}$ can be constructed by a linear combination of the rows in $D$ using the non-negative coefficients $p_{i|t}$ as follows:
\begin{align}
\overline{u}_{i|t} =  p_{i|t}^{\top}D.\label{eqn:s2p_dmp}
\end{align}
Here, $d_{\mathrm{dict}}$ denotes the number of rows in the dictionary $D$ and $d_\mathrm{user}$ denotes the number of hidden units to represent user preferences. The dictionary $D$ can be regarded as a collection of \textit{basis} which is capable of modeling the changes of user preference vector according to his/her sequential behavior. Then, the combined preference vector of user $i$ at time $t$ can be modelled as follows:
\begin{align}
u_{i|t}= u_i^{\ast} + \overline{u}_{i|t},\label{eqn:m_infer}
\end{align}
where $u_i^{\ast}$ is the static user preference vector.

To learn the optimal dictionary $D$ for modeling user dynamic preferences, we can define a optimization objective as follows:
\begin{align}
\mathcal{D}(r_{i}, \hat{r}_{i|t} | \theta)
&=\mathcal{D}(r_{i}, (u_i^{\ast} + \bar{u}_{i|t}){V}^\top  | \theta) \nonumber\\
&\iff\mathcal{D}(r_{i} - u_i^{\ast}V^\top, \bar{u}_{i|t}{V}^\top  | \theta) \nonumber\\
&=\mathcal{D}(r_{i} - u_i^{\ast}V^\top, p_{i|t}^{\top}DV^\top  | \theta)\nonumber\\
&=\mathcal{D}(r_{i}^{\mathrm{res}}, p_{i|t}^{\top}DV^\top  | \theta). \label{eqn:dist_nmf}
\end{align}
Here, $r_{i}$ is the rating vector of user $i$ and $\hat{r}_{i|t}$ is the predicted rating vector at time $t$. $\mathcal{D}$ is the distance function between two vectors. Let $r_{i}^{\mathrm{res}} = r_{i} - u_i^{\ast}V^\top$ and $D' = DV^\top$ and $\mathcal{D}$ be the F-norm. The above Eqn.~\ref{eqn:dist_nmf} means that minimizing the discrepancy between the true ratings and predicted ratings is equivalent to minimizing the discrepancy between the residual error (true rating minus predicted rating from static user preferences) and the dynamic rating (computed from dynamic user preferences). Based on this idea, the above optimization objective can be reformulated as follows:
\begin{align}
\min_{p_{i|t}, D'} || r_{i}^{\mathrm{res}} - p_{i|t}D'||~~\mathrm{s.t.,}~~p_{i|t} \ge{0}, ~~\sum_{k} p_{i|t}(k) = 1. \nonumber
\label{eqn:ls_nmf}
\end{align}
Therefore, modeling the dynamic user preferences can be formulated as a supervised dictionary learning problem, in which the dictionary D is randomly initialized and then optimized by gradient-based learning techniques.
Note that the above Eqn.~\ref{eqn:dist_nmf} only illustrates our main idea that the goal of the dictionary learning is to minimize the residual error of the predictions. More specifically, we use an end-to-end training for the whole model instead of training each part of the model individually. The final loss functions are presented in Section~\ref{sec:loss}.

In the above dictionary learning problem, the posterior distribution $p_{i|t}$ should not be obtained via point estimations which are not feasible in test time. Therefore, different from standard dictionary learning problem~\cite{mairal2009online}, we propose to learn a state function $\phi$, which allows us to accurately obtain $p_{i|t}$ from the user rating vector as follows:
\begin{align}
p_{i|t}=\phi(r_{i_{\le{t}}}).
\end{align}
$r_{i_{\le t}}$ denotes the complete rating history of user $i$ before time $t$. Note that the learning of $\phi$ is challenging because (1) it should accurately capture the sequential dependencies within the rating sequence of each user and (2) it needs to embed the sequential dependencies into the same latent space as the downstream score estimator. To this end, we propose to use a RNN-based autoregressive model to capture the sequential dependencies, and then feed the learned information to the latent space via minimizing the loss $\mathcal{D}(r_{i}, \hat{r}_{i|t} | \theta)$. 

\subsubsection{Encoding Sequential Dependencies}
This paper adopts GRU~\cite{cho2014learning} with attention to learn user dynamic preferences. GRU can adaptively capture the sequential dependencies with different time scales which is more suitable to model the users with different rating time scales in recommendation problem. The neural attention mechanism permits learning an adaptive nonlinear weighting function, which allows the more (less) related dependency patterns to make more (less) contributions in the predictions.

Given the historical ratings of a user $\{x_1, \dots, x_T\}$\footnote{When training the model, we use the entire sequence without slicing, so that the length $T$ is variable for each user in a batch.}, we first put them into an embedding layer which outputs continuous vectors, then we feed these vectors to GRU to learn the sequential representations $\{g_1, \dots, g_T\}$. Each hidden state $g_t$ after the $t$-th historical item can be formally described as follows:
\begin{align}
h_{t}= \mathrm{Recurrency}(h_{t-1}, x_t).  \label{eqn:gru_h}\\
\alpha_{t}= \mathrm{Attention}(h_{t}).  \label{eqn:attend_w}\\
\textstyle g_t= \sum_{j=1}^{T}\alpha_{t,j}h_{j}. \label{eqn:attend_g}
\end{align}
In Eqn.~\ref{eqn:gru_h}, GRU is used as a recurrent activation function which we refer to as a \textit{recurrency}. Meanwhile, $h_{t-1}\in\mathbb{R}^{d_\mathrm{GRU}}$ is the ($t$-1)-th hidden state of GRU, and $\alpha_t \in \mathbb{R}^{T}$ is a vector of the attention weights. Eqn.~\ref{eqn:attend_w} describes a content-based multiplicative attention mechanism~\cite{chorowski2015attention} which scores each element in $h$ separately and normalizes the scores using softmax as follows:
\begin{align}
h'_{t}= \tanh(W_{h}h_t +b_h),
\quad e_{t, j}= {h_{j}^{\top}h'_{t}}/{\sqrt{d_{\mathrm{rnn}}}},   \label{eqn:attend_trans}\\
\textstyle \alpha_{t,j} = \exp(e_{t, j})/ \sum_{i=1}^{T} \exp(e_{t, i}). \label{eqn:attend_ws}
\end{align}
Here, we use a weighted mapping $W_{h}\in{\mathbb{R}^{d_{\mathrm{rnn}}\times{d_{\mathrm{rnn}}}}}$
in stead of an identity mapping due to better empirical performance.

\subsubsection{Decoding Dynamic User Preferences}
After encoding the sequential dependencies in a rating sequence, we decode and embed these information into the same latent space of user static preference. More specifically, we learn a multi-layer perceptron (MLP) to approximate the posterior distribution $p_{i|t}$ in Eqn.~\ref{eqn:s2p_dmp}, i.e., the output of the MLP is used as $p_{i|t}$ in  Eqn.~\ref{eqn:s2p_dmp}. Formally, we estimate the posterior distribution $p_{i|t}$ for the $t$-th item rated by user $i$ as follows:
\begin{align}
z_{t} = \mathrm{concat}(h_t, g_t, h_t-g_t, h_t\circ g_t).  \label{eqn:dcd_z}\\
c_{t}=\psi(  W_{z}z_{t} + b_{z} ). \label{eqn:dcd_relu_s}\\
c'_{t, k} = \mathrm{sign}(\mathrm{abs}(c_{t, k})) \circ \exp(c_{t, k}) . \label{eqn:dcd_relu_sparse}\\
\textstyle p_{i|t}(k)= {c'_{t, k}}/{ \sum_j c'_{t, j}}.  \label{eqn:dcd_relu_sm}
\end{align}
Here, the operator $\circ$ is the element-wise product and $h_t\circ{g_t}$ in Eqn.~\ref{eqn:dcd_z} can capture the second-order interactions between the learned sequential dependencies because $g_t$ is a linear combination of $h_{1},\dots, h_T$. In this way, $z_t$ can be more informative due to containing high-order interactions among $h_t$ and can potentially improve model performance~\cite{rendle_tist2012,van2016conditional}. $\psi$ in Eqn.~\ref{eqn:dcd_relu_s} is the activation function, such as \textit{sigmoid}, \textit{tanh} and ReLU~\cite{nair2010rectified}. The term $\mathrm{sign}(\mathrm{abs}(c_{t, k}))$ is the masking function\footnote{Note that $sign(abs(c_{t, k}))$ is a non-differentiable function, of which the gradients are ignored for simplicity. Alternatively, we have also tried the differentiable masking function: $c'_{t, k} = \exp(c_{t, k}) - \exp(0)$, 	but the performance differences are negligible.}, which equals to $0$ if $c_{t, k}$ is $0$ and $1$ otherwise. This design allows for a sparse $p_{i|t}$. Eqn.~\ref{eqn:dcd_relu_sparse} poses a non-negative constraint on $p_{i|t}(k)$ and Eqn.~\ref{eqn:dcd_relu_sm} normalizes the probabilities. This non-negative constraint forces the rows of $D'$ in Eqn.~\ref{eqn:ls_nmf} to combine, not to cancel out, which can yield more interpretable features and improve the downstream prediction performance~\cite{li2016recovery}.

After obtaining $p_{i|t}$, we  compute the final dynamic user preferences by interpolating the dictionary $D$ with the weights $p_{i|t}$ as follows:
\begin{align}
\textstyle \overline{u}_{i|t} = p_{i|t}^{\top}D =  \sum_{k} p_{i|t}(k)D_k. \label{eqn:dyn_user}
\end{align}

\subsection{Estimator}
Similar to the BiasedMF method~\cite{paterek2007improving}, we formulate
the prediction function of our method as follows:
\begin{align}
\hat{r}_{ij|t}(\theta) = b_g + b_i + b_j + (u_i^{\ast} + \bar{u}_{i|t})^{\top}v_j,
\label{eqn:cls_pred}
\end{align}
where $v_j$ is the item embedding, $b_g$ is the average of all ratings,
and $b_i$ and $b_j$ are biases of the user $i$ and item $j$, respectively.
Alternatively, we also tried to replace the inner product in
Eqn.~\ref{eqn:cls_pred} with an MLP $\varphi$ as suggested in~\cite{he2017neural}:
\begin{align}
\hat{r}_{ij|t}(\theta) = b_g + b_i + b_j + \varphi(u_i^{\ast} + \bar{u}_{i|t}, {v_j}).
\label{eqn:mlp_pred}
\end{align}
In this way, we observed accuracy improvements on random data splitting protocol,
but the improvements are negligible on sequential-based data splitting protocol.
Therefore, we still use Eqn.~\ref{eqn:cls_pred} in this paper
due to higher efficiency.

\subsection{The Optimization Objectives for Recommendation}
\label{sec:loss}
There are two main recommendation tasks in the literature:
rating prediction and item ranking. In rating prediction,
we predict how a user will rate an unseen item in the future,
e.g., 1-5 stars. In item ranking, we predict whether a user
will interact with an unseen item in the future, e.g., buy a product or not.
As we can see in previous sections, the proposed Seq2Pref network 
can work properly on both kinds of tasks because we can use user rated items to 
train their dynamic preference. However, the main difference comes from the loss function.
For rating prediction, we can let the Seq2Pref network to predict the rating of 
next rated item in a sequence and backpropagate the error to update the parameters.
For item ranking, the prediction errors on both rated and unrated items should 
be considered to update the parameters. To this end,  different optimization 
objectives should be defined for the two tasks.

Let $\theta$ be the model parameters, $\hat{r}_{ij|t}(\theta)$ be
the predicted score of user $i$ on item $j$ given $\theta$ and
$\mathcal{I}_\mathrm{train}$ be the training set, we adopt the
popular mean square loss for rating prediction task~\cite{koren2008factorization,paterek2007improving}
as follows:
\begin{align}
\min_{\theta} \sum_{(i,j,t)\in{\mathcal{I}_\mathrm{train}}}
\left( r_{ij} - \hat{r}_{ij|t}(\theta)\right)^{2} + R(\theta).
\label{eqn:obj_func}
\end{align}
$R(\theta)$ is a regularization term to prevent overfitting.

For item ranking task, we assume all unrated items are negative 
following many existing works~\cite{Hu08,he2016fast} and we sample fraction of 
negative examples for faster training~\cite{chen2017sampling}.
More specifically, we adopt the popular weighted mean square loss~\cite{Hu08,he2016fast} as follows:
\begin{align}
\min_{\theta} \sum_{(i,j,t)\in{\mathcal{I}_\mathrm{train}}}
w_{ij}\left( r_{ij} - \hat{r}_{ij|t}(\theta)\right)^{2} + R(\theta).
\label{eqn:obj_func_rank}
\end{align}
Here, $w_{ij}$ denotes the weight of user $i$ on item $j$,
in which $w_{ij}$ is large for the true positive ratings and
small for the negative ratings to address the implicit feedback
issue in item ranking task~\cite{Hu08}.
Therefore, S2PNM can be effectively trained with the
back-propagation algorithm via stochastic gradient descent.
More training details can be found in the experiment section.

\subsection{Parallel Training}
The proposed S2PNM method will suffer from efficiency issue on large datasets
similar to many existing neural network-based methods~\cite{he2017neural,HidasiICLR16,rrn2017,baral2018close}.
However, we can leverage multiple GPUs to largely reduce the training time.
For the proposed Seq2Pref network, we can use the mini-batch parallel
tricks~\cite{HidasiICLR16} to address length variation issue
(e.g., the sequence length may range from 2 to 17770 on the Netflix Prize dataset),
in pursuit of higher scalability. More specifically, we use a sliding window
over the sequence and put the windowed fragments, referred to as mini-batch,
next to each other. Then in the training process, if any of the mini-batches
finishes, the next available mini-batch is filled in.
We observed that it takes about 4 minutes per iteration using one GTX
1080Ti GPU on the Netflix prize dataset, and the time reduces to 3 minutes
with two GPUs.

\begin{table}[t!]
	\caption{Statistics of the evaluation datasets.}
	\centering
		\begin{tabular}{ l | c | c | c | l }\hline
			Datasets & \#Users & \#Items  &  \#Interactions & Density
			\\\hline\hline
			Instant Video & $1,372$ & $7,957$ & $23,181$ & $0.43\%$
			\\\hline
			Baby Care & $5,057$ & $10,420$ & $77,787$ & $0.15\%$
			\\\hline
			Netflix & $480,188$ & $17,769$ & $100,462,737$ & $1.18$ \%
			\\\hline
	\end{tabular}
	\label{tbl:stats_data}
\end{table}
\section{Experiments}
\label{sec:exp_eval}
In this section, we first empirically study the performance of static and dynamic user preferences with varying hyper-parameters. Then, we compare the accuracy of the proposed method on both rating and ranking tasks, compared against state-of-the-art methods. Qualitative analysis on the Netflix prize dataset demonstrate that S2PNM can indeed capture the user preference drifts over time.

\subsection{Experimental Setup}
\label{sec:exp_setup}

\subsubsection{Datasets}
Two widely adopted real-world datasets are used in the experiments:
(1) Netflix Prize\footnote{https://www.netflixprize.com/}~\cite{Netflix07}
for rating prediction task, and (2) Amazon dataset\footnote{http://jmcauley.ucsd.edu/data/amazon/}~\cite{he2016ups} for
item ranking task. For the Netflix dataset, we split it into training and test sets
by chronological order to simulate the real scenarios, and we set the ratio
of training and test sets as 9:1. For the Amazon dataset, we use two product
categories including Instant Video and Baby Care. To achieve sequential
recommendation, we select users with at least 10 purchasing records in the
experiments. Each user's purchasing history is ordered by purchasing time,
and the first 70\% items of each user are used for training while the remaining
items are used for test. The statistics of the final datasets are shown in Table~\ref{tbl:stats_data}.

\subsubsection{Training Details}
\label{sect:train_details_s2pmn}
To train S2PNM, we adopt the adaptive learning rate algorithm --
Adam~\cite{kingma2015adam} with $\beta_1=0.9$, $\beta_2=0.98$ and
$\epsilon=10^{-9}$. We also decay the learning rate over 5 full
data passes with a rate of $0.9$. Meanwhile, all the weight matrices are initialized from a Glorot uniform distribution~\cite{glorot2010understanding},
and recurrent weights are furthermore orthogonalized. 
We also employ dropout~\cite{srivastava2014dropout} as regularization during training RNNs with a dropout rate of $p_{\mathrm{drop}}=0.02$, and $\ell_2$-norm as regularization to penalize the user and item embeddings. 
The historical items of each user are batched together with a batch size of 16. We train S2PNM with 20 epochs and measure the performance after each epoch. Then, we report the results on the test set using the best performing model on the validation set. We found that S2PNM is quite robust to hyper-parameters, therefore we adopt the same hyper-parameter setting across both datasets.
In addition, pretraining the static user/item preferences can achieve higher accuracy and faster convergence speed. Therefore, in the experiments, we use the BiasedMF method~\cite{paterek2007improving} to initialize the static user/item preference vectors. Training details of the compared methods can be found below.

We tune the learning rate $lr \in$ \{0.001, 0.002, 0.005, 0.01, 0.02, 0.05, 0.1$\}$, the regularization strength $\lambda \in \{ 0.001, 0.01, 0.1\}$ and the latent factor dimension $d\in \{20, 50, 100, 150, 200, 300\}$ via grid search for all compared methods. Other implementation details are as follows:
\begin{itemize}
	\item \textbf{SVD++}~\cite{koren2008factorization} is one of the most popular hybrid collaborative filtering based approach. We use the cpp implementation in GraphChi~\footnote{https://github.com/GraphChi/graphchi-cpp}~\cite{kyrola12graphchi}, and we have found that the optimal results on both MovieLens 10M and Netflix data can be achieved when we use factor size $200$, learning rate $0.002$ with decay rate $0.99$ and regularizer $0.01$.
	\item \textbf{GLOMA}~\cite{chen2017gloma} achieves robust results on many benchmarks by enhancing local models based on submatrix with a unified global model. We use StableMA~\footnote{https://github.com/ldscc/StableMA} provided by the authors which is implemented in Java, and adopt the default setting in~\cite{chen2017gloma} which shows the best RMSEs in both idealistic and realistic scenarios, i.e., learning rate $0.0008$, regularization $0.06$, and the latent factor dimension $300$. Notably, we introduce biases into GLOMA following the same idea in BiasMF~\cite{koren2009matrix} for improved accuracy, particularly in realistic scenario it helps improve the model performance from 0.96302 to 0.89326 on Netflix data.
	\item \textbf{MRMA}~\cite{mrma} is so far among the best ensemble based recommendation algorithm. We use the program provided by the authors, and the default setting in~\cite{mrma} produces the best RMSEs in both scenarios. That is factor size ranging in $\{10,20,50,100,150,200,250,300\}$, learning rate $0.001$ and regularizer $0.02$. As the same with GLOMA, we modify MRMA in a fashion of BiasMF, which reduces the RMSE on Netflix from 0.94780 to 0.89150 in realistic scenario.
	\item \textbf{TimeSVD++}~\cite{koren2010collaborative} is one of the most successful models which are able to capture dynamic nature of the recommendation data. We test the implementation in LibRec~\cite{guo2015librec}\footnote{https://github.com/guoguibing/librec}, in addition to GraphChi \cite{kyrola12graphchi} of which the implementation is slightly different from the original paper (See lines 162 - 167 in timesvdpp.cpp for more details). This explains why the result of LibRec i.e., RMSE 0.90 on Netflix is better than that from GraphChi i.e., RMSE 0.92 in realistic scenario, where noticeably our results for GraphChi are comparable to~\cite{rrn2017}. While for LibRec it takes $\ge$24 hours per iteration on Netflix data, by rewriting the data structure to store each user's data and refining the model update procedure we reduce it to nearly 2 mins per iteration. Although we tuned this model very carefully, it is still worse than SVD++, and the best results are observed  by using factor size $200$, learning rate $0.002$ for latent factors and $0.00003$ for bias parameters, regularizer $0.01$.
	\item \textbf{AutoRec}~\cite{autorec} is among the best neural network models so far in terms of rating prediction. We use the program provided by the authors~\footnote{https://github.com/mesuvash/NNRec}, which indeed reproduces the results shown in the paper. However, we fail in performing it on Netflix data due to the requirement of memory more than 150 GBs. Therefore, we have to implement it in modern deep learning platform - Tensorflow, whereby we achieve RMSE 0.78056 on MovieLens 10M which is much better than 0.78463 produced by the authors' code. More specifically, we adopt the latent state dimension as $500$ to yield the best performance and use Adam~\cite{kingma2015adam} with batch size $512$, learning rate $0.0005$ with decayed rate $0.9$ every $10$ full data passes  to train the model.
	\item \textbf{NADE}~\cite{zheng2016neural} learns the ordinal nature of the user preference and achieves the best results among all baselines. We use the program provide by the authors~\footnote{https://github.com/Ian09/CF-NADE} to evaluate U-CF-NADE-S, and meanwhile we also test the program based on Chainer~\footnote{https://github.com/dsanno/chainer-cf-nade} which produces similar results on MovieLens 10M. The Chainer version requires much less memory as a result of storing data in sparse matrix, in contrast to dense matrix in authors' code. For computational time, it takes 15 mins per iteration but requires 500 iterations to converge.  After fine-tuning the model with 8 GTX-1080Ti, we have found that the optimal setting varies in different datasets, but the configurations with hidden unit size of $500$ in original paper always produce the best results.
	\item \textbf{RRN}~\cite{rrn2017} is one of the most closely related works which takes temporal dynamics into consideration, and offers excellent prediction accuracy. We use the program provided by the authors, which is relied on a late-2016 version of MXNET. We rewrite part of the code to make it work with the dependency mxnet-cu80 of version 0.9.5. Different from its original paper, we use $200$-dimensional stationary factors, extra $120$-dimensional dynamic factors, and a single-layer LSTM with $100$-hidden neurons, pursuing better results. Notably, we use BiasMF to initiate RRN model instead of PMF~\cite{mnih2007probabilistic} and Autorec~\cite{autorec} for improved performance -- RMSE reducing from 0.91823 to 0.89014.
	
	\item \textbf{BPR}~\cite{rendle2009bpr} is the most famous pairwise matrix approximation based model for ranking prediction. We use the implementation in LibRec~\cite{guo2015librec}. By experiments, we found large learning rates lead to accelerated convergence rate so that we chose  $0.25$ on Baby Care dataset. In addition, we also found that decayed learning rates can significantly improve the prediction accuracies, perhaps we decayed the learning rate by 0.9 every epoch. After grid search, we use the factor size $128$ and the regularizer $0.01$ on Baby Care dataset.
	\item \textbf{eALS}~\cite{he2016fast}  is one of the most successful pointwise recommendation approaches. We use the implementation in LibRec~\cite{guo2015librec}. We select factor size over $\{16, 32, \dots, 256\}$ and regularization parameter over $\{0.1, 0.2,\dots,1.0\}$. To be specific, we use the factor size $64$ for all datasets, and $c_0$=$256$, learning rate $0.01$ and regularizer $0.1$ on Baby Care dataset.
	\item \textbf{NeuMF}~\cite{he2017neural} is among the best neural models for ranking prediction. We use the program provided by the authors \footnote{https://github.com/hexiangnan/neural\_collaborative\_filtering} , where we search by grid the learning rate over \{\textit{1e-1}, \textit{5e-2},$\dots$,\textit{1e-4}\}, batch size over \{$256$, $512$, $\dots$, $4096$\}. The batch size of $2048$ is the optimal for all datasets. And the best performance is achieved when we use learning rate \textit{1e-3}.

	\item \textbf{RUM}~\cite{chen2018sequential} is one of state-of-the-art sequential recommendation models based on memory networks, and also most closely related to our work. We note that the comparisons to the classic FPMC~\cite{RendleWWW10} and DREAM~\cite{YuSIGIR16} are omit, since RUM excels them by a large margin. We search by grid the embedding dimension $D$ in the range of \{32,  64, $\dots$, 256\}. By experiments, we found the number of memory slots $20$ and embedding size $64$ produce the best results.
	\item \textbf{SHAN}~\cite{ying2018sequential} introduces hierarchical attention networks for temporal modelling and achieves state-of-the-art results on many benchmarks. We search by grid the embedding size over \{$50$, $100$, $\dots$, $300$\} and regularization strength over \{$0.001$, $0.01$, $\dots$, $1$\}. Experiments show that embedding size $300$ and regularizer $0.01$ for user/item factors and regularizer $1$ for transformation matrix yield the best results.
	\item \textbf{SASRec}~\cite{kang2018self} adapts the transformer~\cite{vaswani2017attention} to recommender systems and is among the best sequential recommendation algorithm. We use the implementation provided by the authors~\footnote{https://github.com/kang205/SASRec}, and select the embedding size over \{$50$, $100$, $\dots$, $300$\} and the number of blocks up to $4$. By experiments, the best results can be obtained by using $250$-dimensional embedding size and $4$ blocks with dropout rate $0.2$.
\end{itemize}

\subsubsection{Evaluation Metrics}
We study the performance of the proposed S2PNM model for both rating prediction task and item ranking task by using the popular evaluation metrics in the literature of recommender systems.
For the rating prediction task, we use root mean square error (RMSE) which is defined as $\mathrm{RMSE} = \sqrt{ 1/|\mathcal{I}_\mathrm{test}|   \sum_{(i,j,t)\in{\mathcal{I}_\mathrm{test}}} (r_{ij} - \hat{r}_{ij|t})^2}$ where $\mathcal{I}_\mathrm{test}$ stands for the set of test examples.

For the item top-k ranking task, we evaluate the proposed S2PNM model by using Precision@$k$, Hit-Rate@k (HR@$k$) and Normalized Discounted Cumulative Gain@k (NDCG@$k$):
1) $\mathrm{Precision@k} =\frac{1}{|\mathcal{I}_\mathrm{test}|} \sum_{i \in\mathcal{I}_\mathrm{test}}  |R_i \cap T_i|/|R_i|$ where $R_i$ is defined as the $k$-size generated recommendation list for user $i$ and $T_i$ is the grand-truth;
2) $\mathrm{HR@k} = \frac{1}{|\mathcal{I}_\mathrm{test}|} \sum_{i \in\mathcal{I}_\mathrm{test}} \mathbf{1}|R_i \cap T_i|$, in which $\mathbf{1}|x|$ is an indicator function whose value is $1$ when $x>0$ and $0$ otherwise;
3) $\mathrm{NDCG@k} = \frac{1}{|\mathcal{I}_\mathrm{test}|} \sum_{i \in\mathcal{I}_\mathrm{test}}  \mathrm{DCG@k}_i / \mathrm{iDCG@k}_i$, in which $\mathrm{DCG_i@k} = \sum_{j=1}^{k} (\mathbf{1}|R_i^j \cap T_i| - 1)/ \log_2 ( j + 1 )$ and $\mathrm{iDCG_i@k}$ is a normalized constant which is the maximum possible value of $\mathrm{DCG_i@k}$.

\begin{figure*}[tbh!]
	\centering
	\includegraphics[width=.325\textwidth]{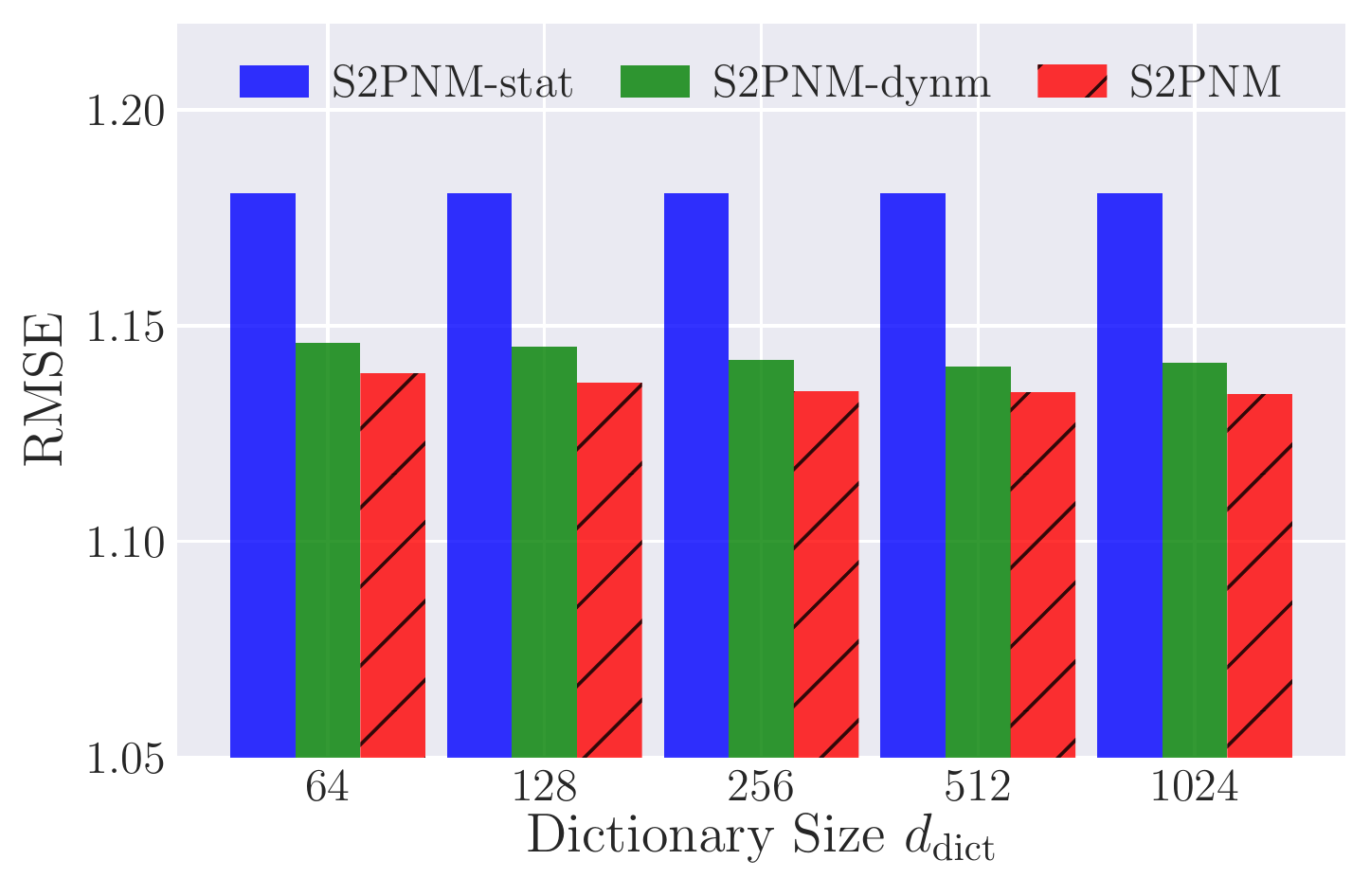}\hfill
	\includegraphics[width=.325\textwidth]{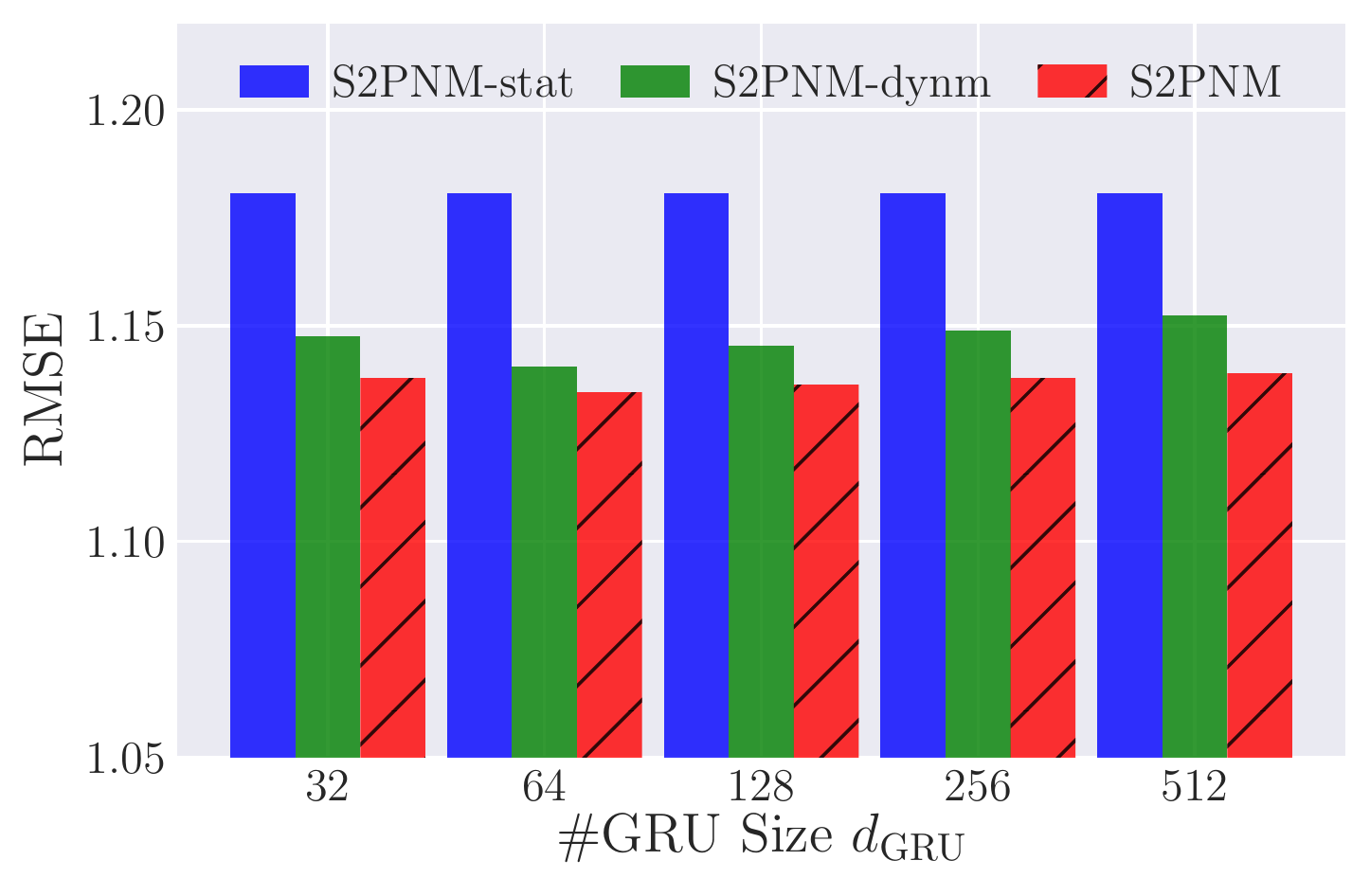}\hfill
	\includegraphics[width=.325\textwidth]{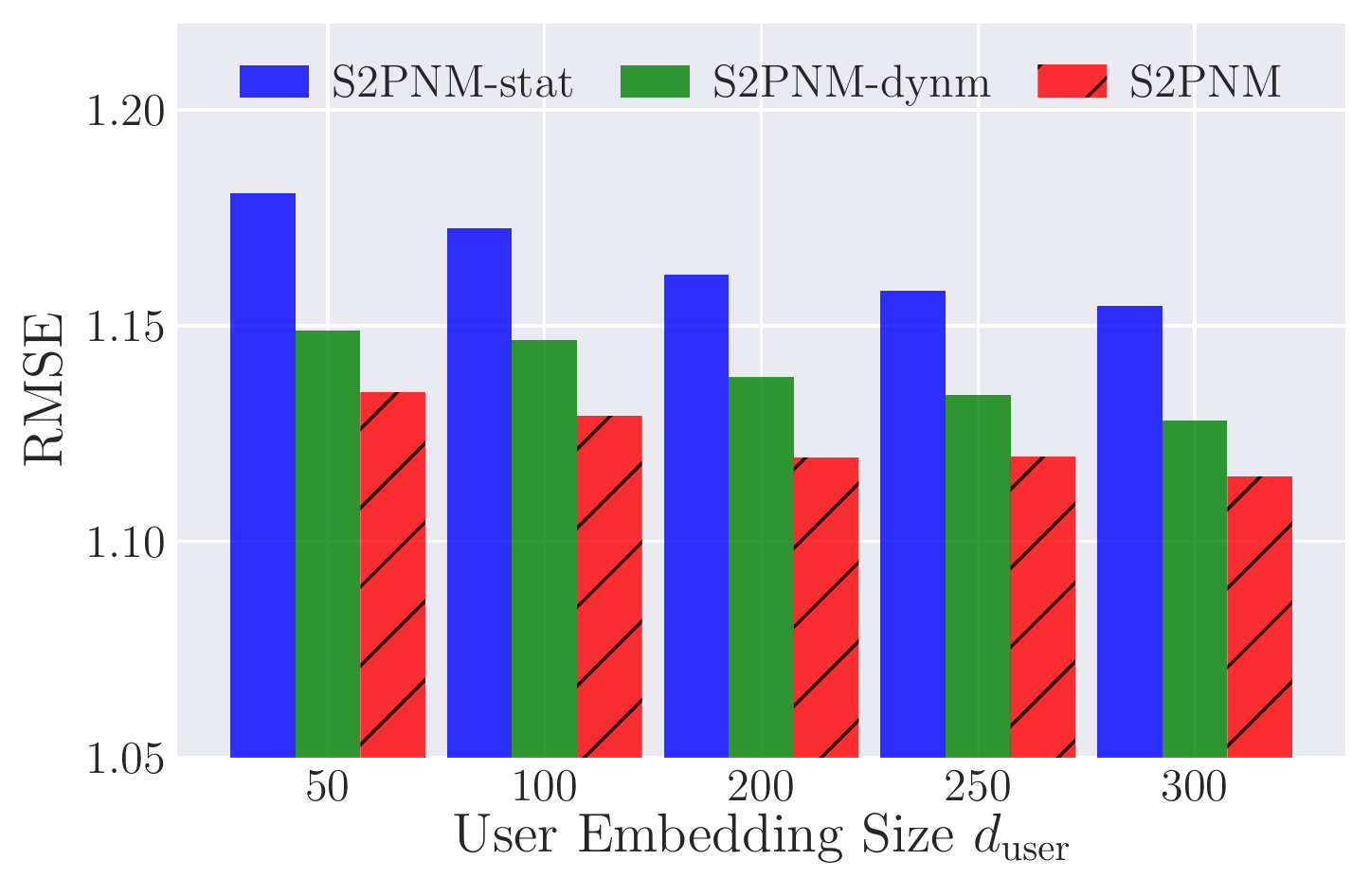}
	\caption{RMSE of S2PNM with static user preferences (S2PNM-stat), S2PNM with dynamic user preferences (S2PNM-dynm), and S2PNM with both static and dynamic preferences (S2PNM), with varying dictionary size $d_\mathrm{dict}$ in $\{64,128,256,512,1024\}$ (left), GRU size $d_\mathrm{GRU}$ varying in $\{32,64,128,256,512\}$ (middle) and user embedding size $d_\mathrm{user}$ (right) varying in $\{50,100,200,250,300\}$ on the Amazon Instant Video.The lower value means the better performance. Note that the accuracy of S2PNM slightly increases when we increase the dictionary size and GRU size.\label{fig:rmse_aiv}}
\end{figure*}
\begin{figure*}[tbh!]
	\includegraphics[width=.325\textwidth]{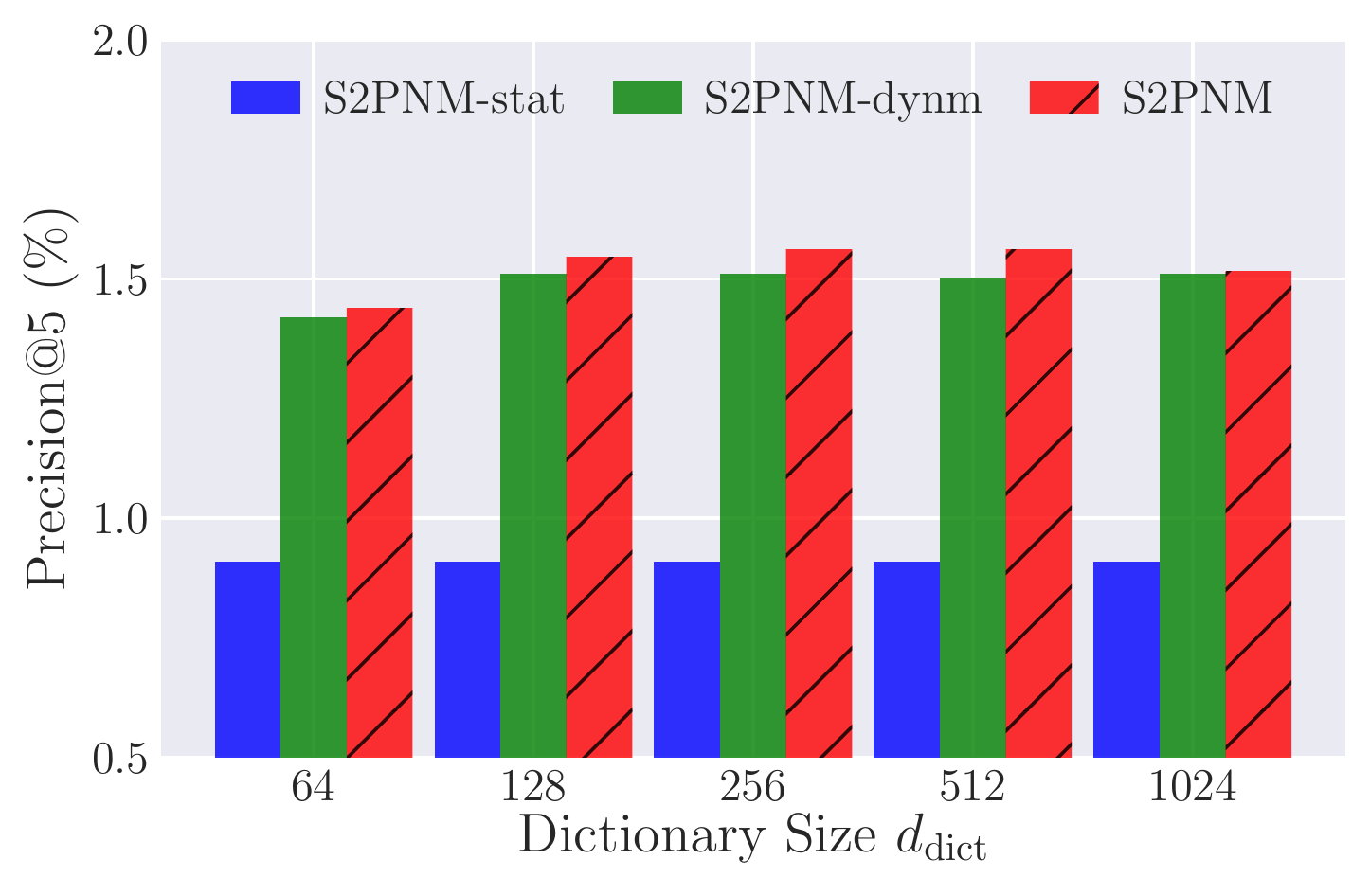}
	\includegraphics[width=.325\textwidth]{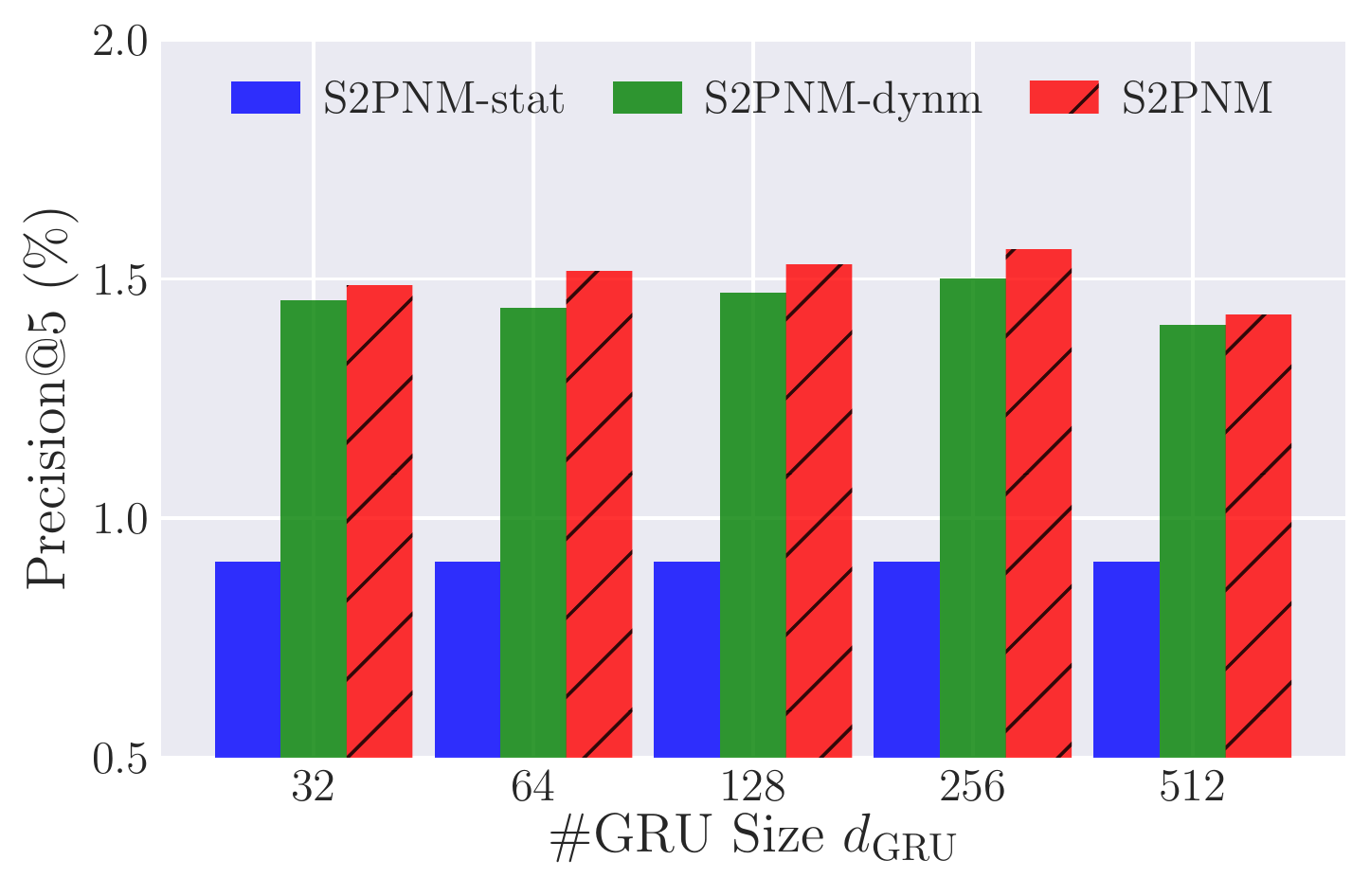}
	\includegraphics[width=.325\textwidth]{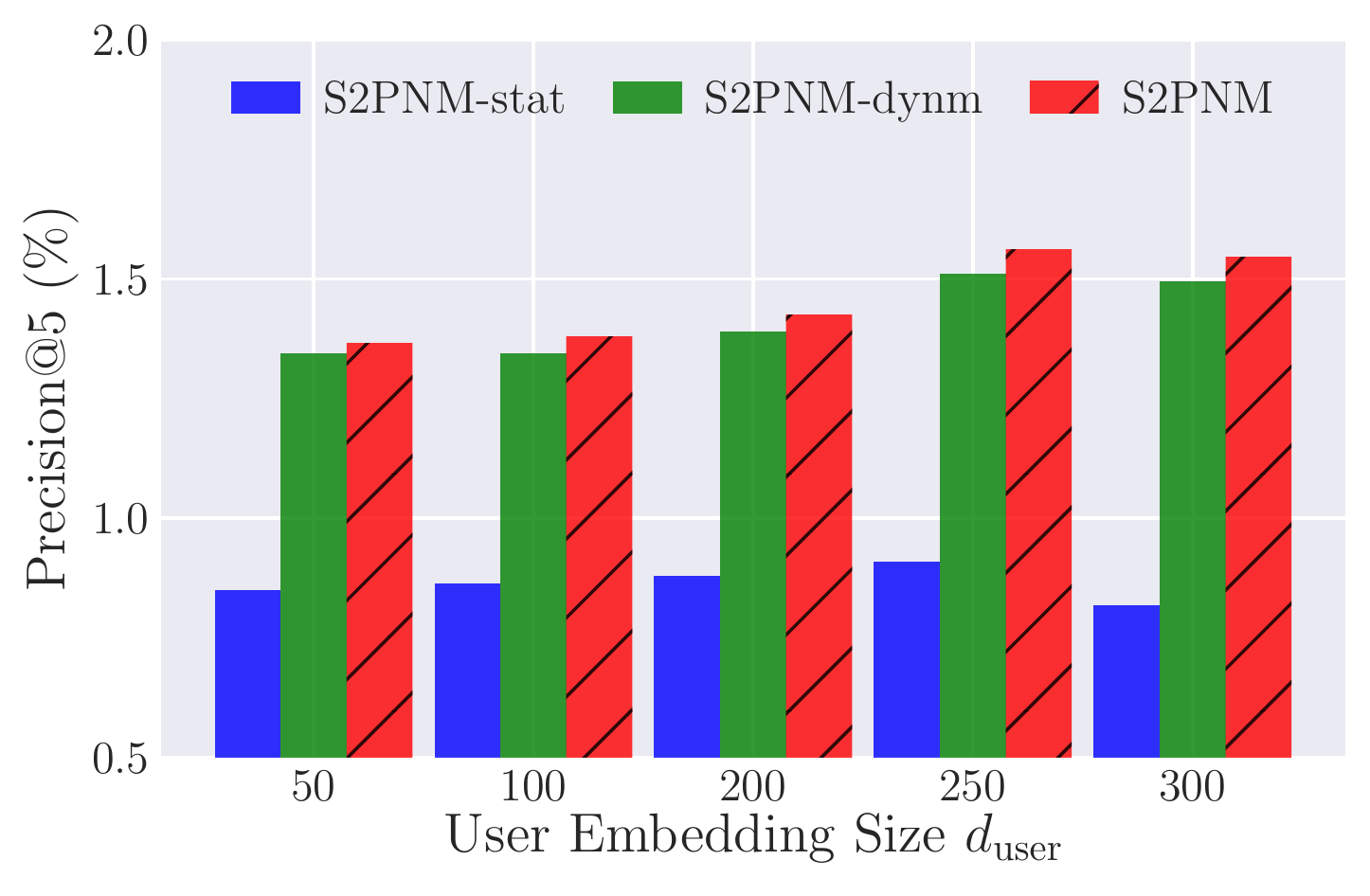}
	\caption{\label{fig:p_aiv}Precision@5 of S2PNM with static user preferences (S2PNM-stat), S2PNM with dynamic user preferences (S2PNM-dynm), and S2PNM with both static and dynamic preferences (S2PNM), with varying dictionary size $d_\mathrm{dict}$ in $\{64,128,256,512,1024\}$ (left), GRU size $d_\mathrm{GRU}$ varying in $\{32,64,128,256,512\}$ (middle) and user embedding size $d_\mathrm{user}$ (right) varying in $\{50,100,200,250,300\}$ on the Amazon Instant Video. The greater value means the better performance.}
\end{figure*}
\begin{figure*}[tbh!]
	\includegraphics[width=.325\textwidth]{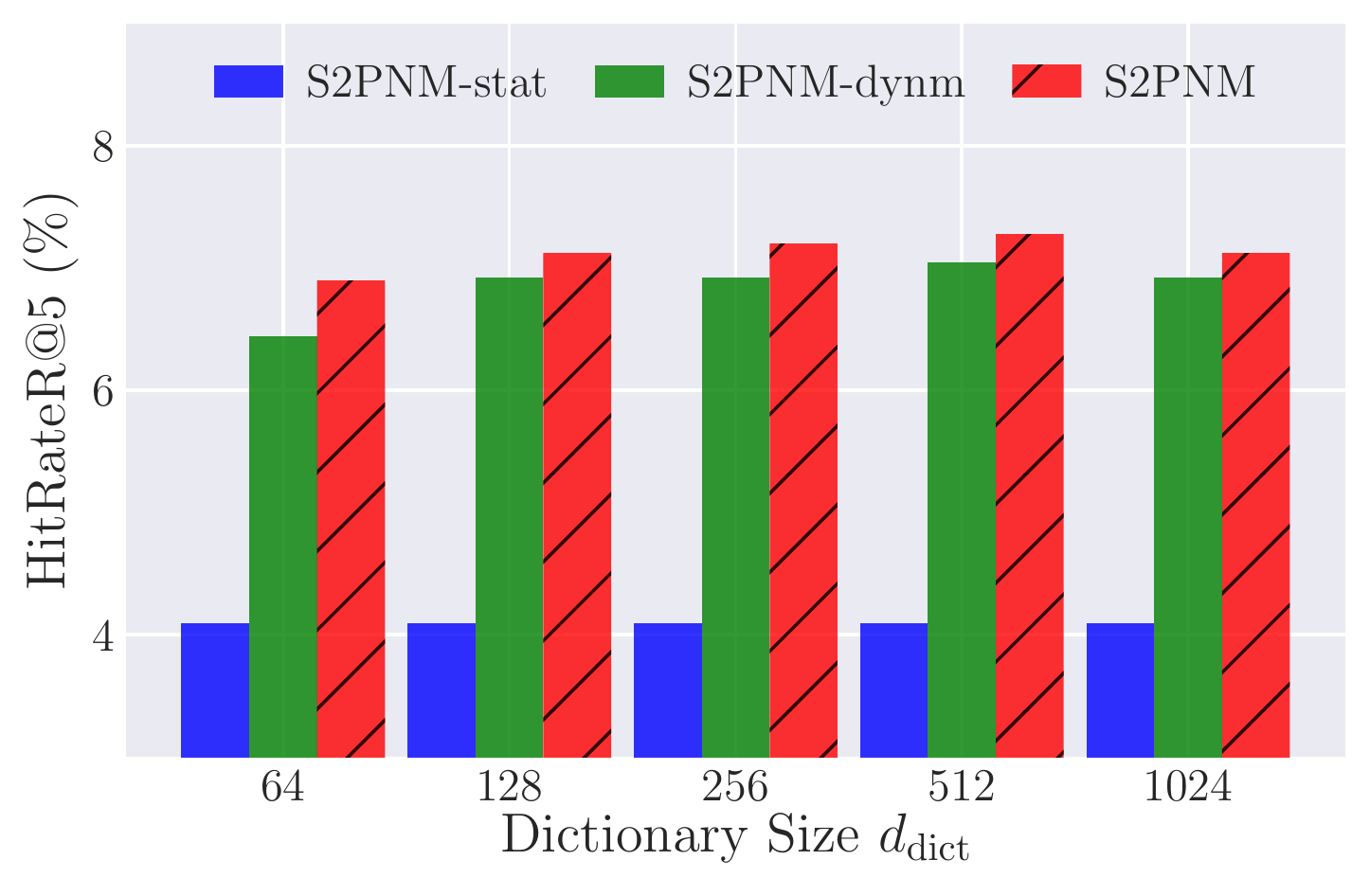}
	\includegraphics[width=.325\textwidth]{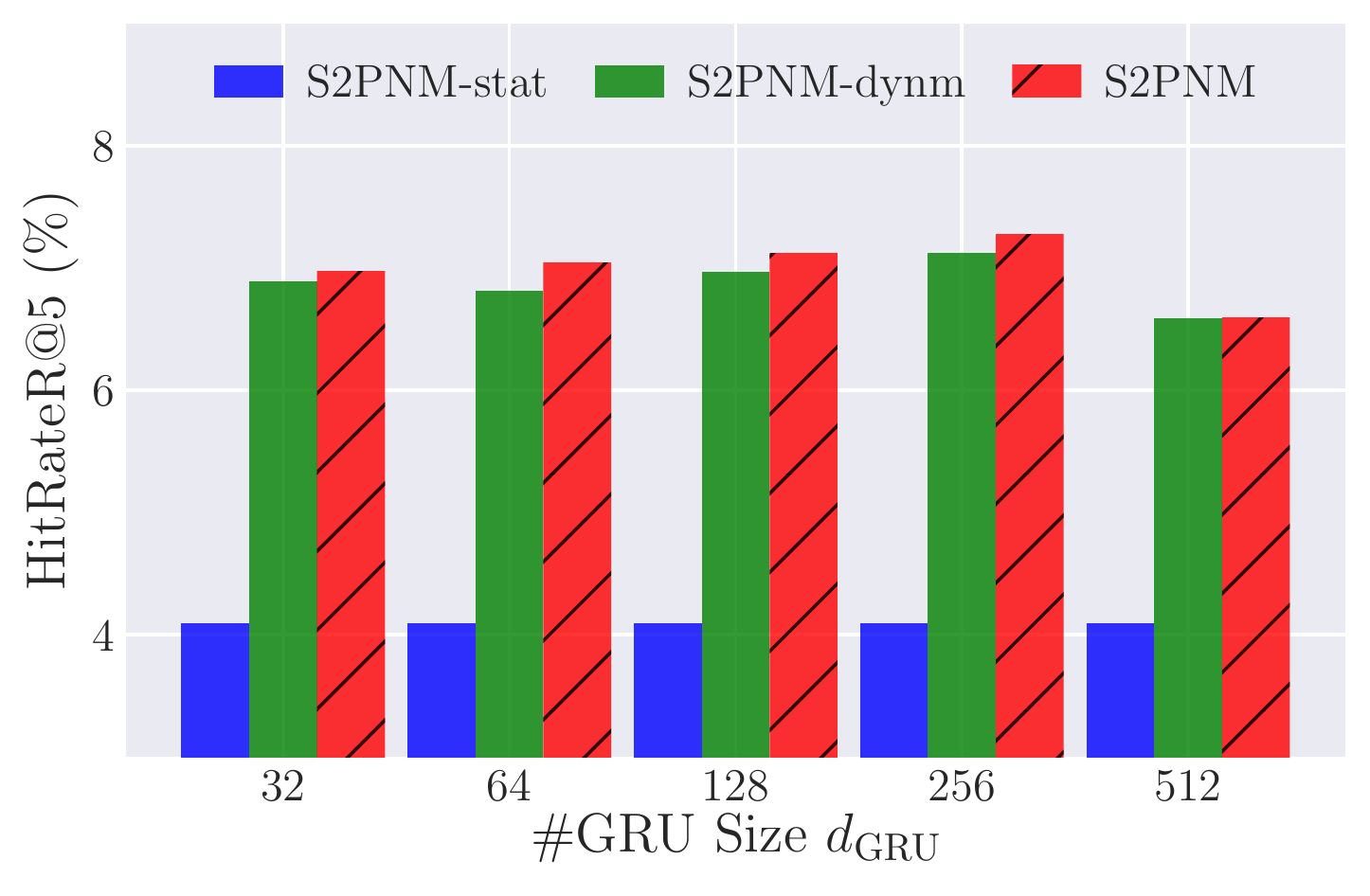}
	\includegraphics[width=.325\textwidth]{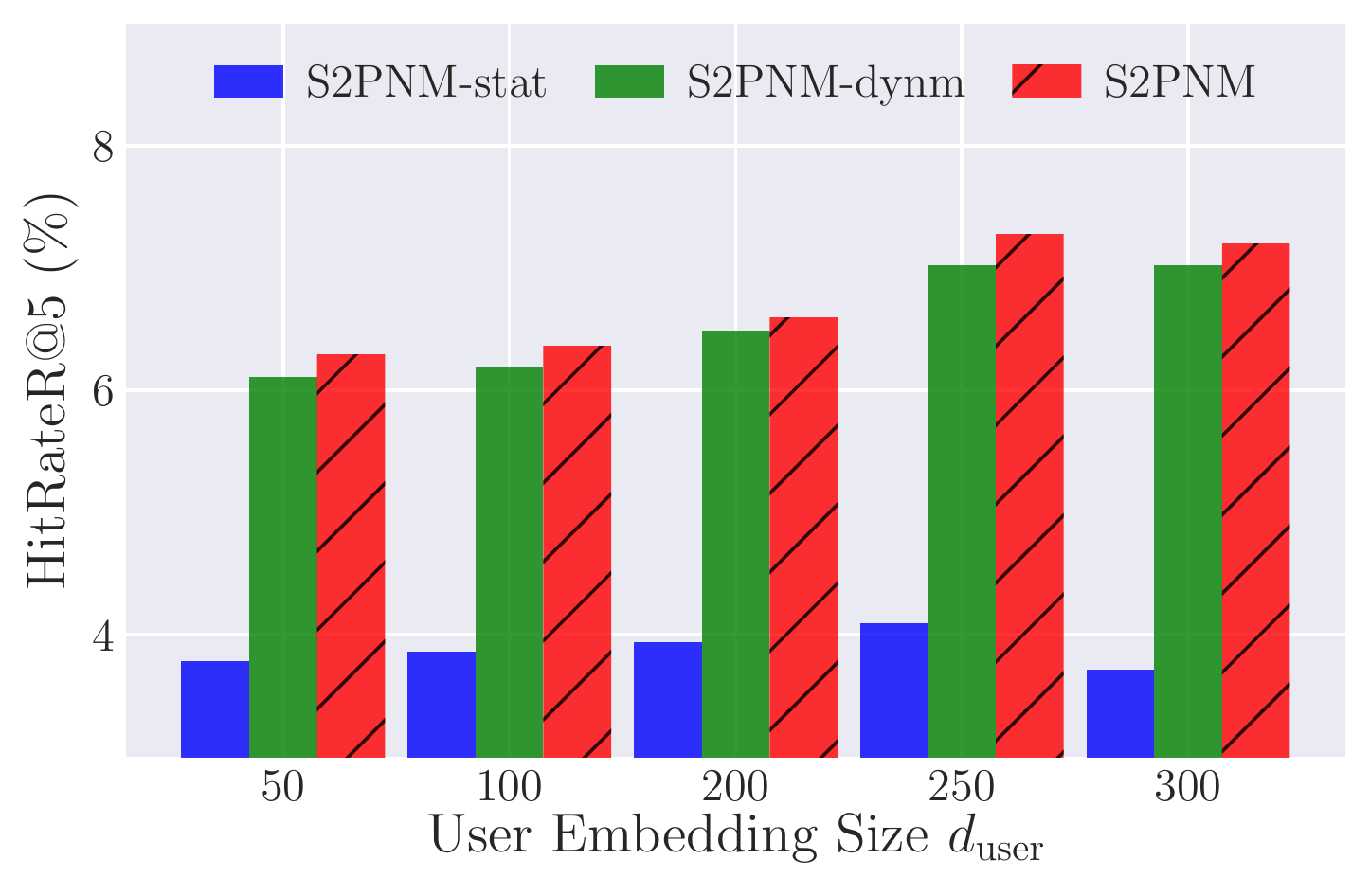}
	\caption{\label{fig:h_aiv}Hit-rate@5 of S2PNM with static user preferences (S2PNM-stat), S2PNM with dynamic user preferences (S2PNM-dynm), and S2PNM with both static and dynamic preferences (S2PNM), with varying dictionary size $d_\mathrm{dict}$ in $\{64,128,256,512,1024\}$ (left), GRU size $d_\mathrm{GRU}$ varying in $\{32,64,128,256,512\}$ (middle) and user embedding size $d_\mathrm{user}$ (right) varying in $\{50,100,200,250,300\}$ on the Amazon Instant Video.The greater value means the better performance.}
\end{figure*}
\begin{figure*}[tbh!]
	\includegraphics[width=.325\textwidth]{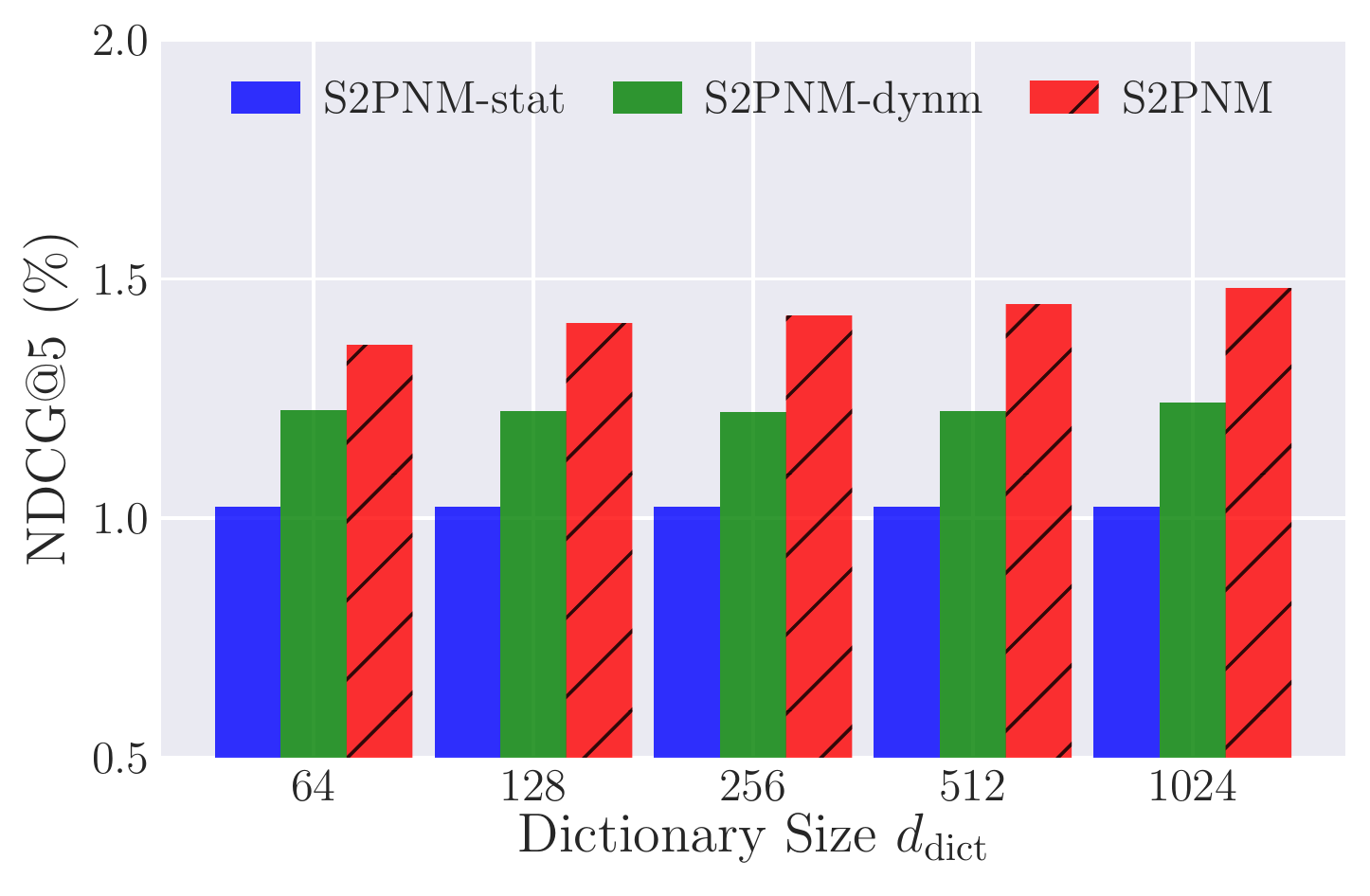}
	\includegraphics[width=.325\textwidth]{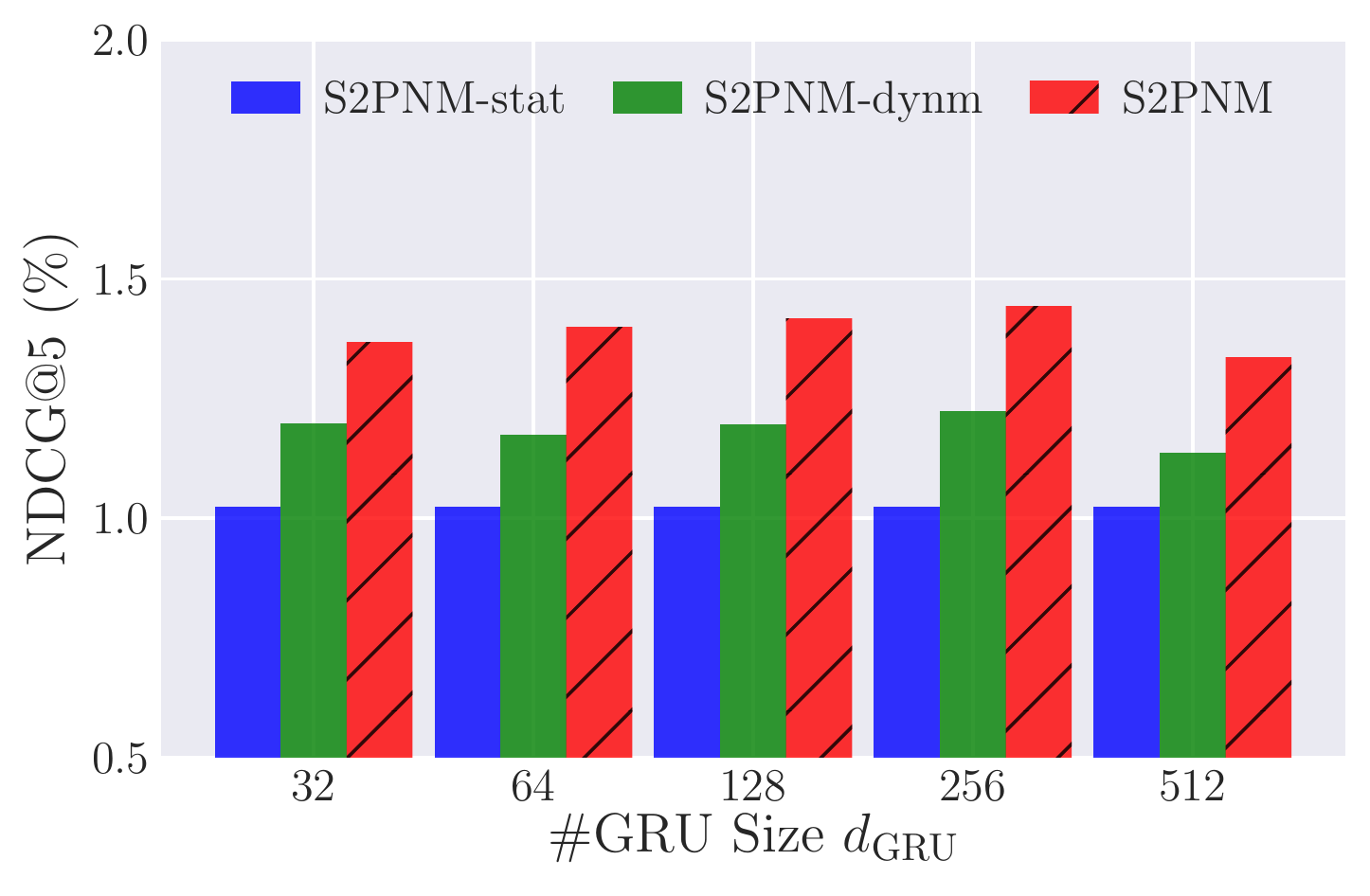}
	\includegraphics[width=.325\textwidth]{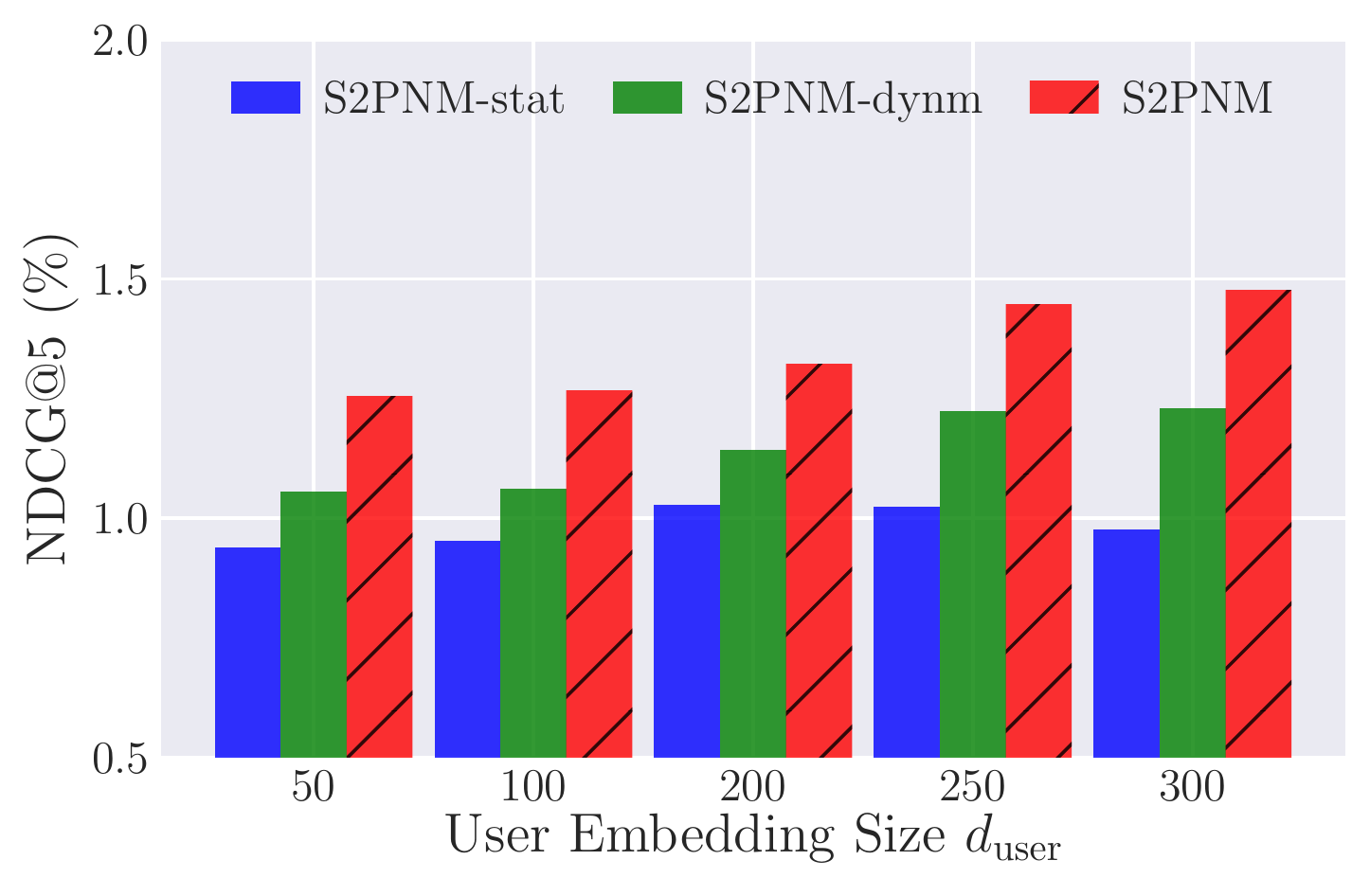}
	\caption{\label{fig:n_aiv}NDCG@5 of S2PNM with static user preferences (S2PNM-stat), S2PNM with dynamic user preferences (S2PNM-dynm), and S2PNM with both static and dynamic preferences (S2PNM), with varying dictionary size $d_\mathrm{dict}$ in $\{64,128,256,512,1024\}$ (left), GRU size $d_\mathrm{GRU}$ varying in $\{32,64,128,256,512\}$ (middle) and user embedding size $d_\mathrm{user}$ (right) varying in $\{50,100,200,250,300\}$ on the Amazon Instant Video.The greater value means the better performance.}
\end{figure*}

\subsection{Sensitivity Analysis}
This study uses the Amazon Instant Video dataset to analyze
the importance of each component in S2PNM, and show how S2PMN performs
with different hyper-parameters.

\subsubsection{Effects of static and dynamic preferences}
S2PNM-stat and S2PNM-dynm denote the S2PNM model with solely using static user preferences and dynamic users preferences, respectively. Figure.~\ref{fig:rmse_aiv} - \ref{fig:n_aiv} show that S2PNM-dynm outperforms S2PNM-stat in all cases, and S2PNM (with both static and dynamic user preferences) achieves much higher accuracy than both S2PNM-dynm and S2PNM-stat. These results confirm that (1) S2PNM is effective to capture user dynamic preferences and (2) static user preferences are also necessary for predicting user future ratings without which the performance will be suboptimal. In addition, we can see that S2PNM-dynm and S2PNM have comparable results in terms of Precision@5 and HR@5, whereas S2PNM significantly outperforms S2PNM-dynm in NDCG@5. This indicates that users' static preferences can help to put right recommendations in higher positions.

\subsubsection{Effects of dictionary size}
The leftmost figures of Fig.~\ref{fig:rmse_aiv} - \ref{fig:n_aiv} show the accuracy with different dictionary sizes, i.e., $d_\mathrm{dict}$ varies in $\{64,128,256,512,1024\}$. In Fig.~\ref{fig:rmse_aiv} (left), we set user embedding size $d_\mathrm{user}$ = 50 and the number of hidden units in GRU $d_\mathrm{GRU}$ = 32. In Fig.~\ref{fig:p_aiv} (left), we set $d_\mathrm{user}$ = 250 and $d_\mathrm{GRU}$ = 256. As shown in the results, increasing the dictionary size can increase accuracy when $d_\mathrm{dict}$ varies in  $\{64,128,256,512,1024\}$. This makes sense because a large dictionary can increase the capacity of S2PNM. Therefore, we choose $d_\mathrm{dict}$ = 1024 for the accuracy comparison.

\subsubsection{Effects of GRU size}
The middle figures of Fig.~\ref{fig:rmse_aiv} - \ref{fig:n_aiv} show
the recommendation accuracy with varying number of hidden units in GRU
($d_\mathrm{GRU}$). In Fig.~\ref{fig:rmse_aiv} (middle), we set $d_{\mathrm{user}} = 50$
and $d_\mathrm{dict} = 64$. In Fig.~\ref{fig:p_aiv} (middle),
we set $d_{\mathrm{user}} = 250$ and $d_\mathrm{dict} = 512$.
As shown in the results, increasing the GRU size can improve
the model performance when $d_\mathrm{GRU} \le 256$. However, all ranking
metrics dropped when $d_\mathrm{GRU} = 512$, which suggests that overfitting
happens with $d_\mathrm{GRU} = 512$. Therefore, we choose $d_\mathrm{GRU} = 256$
for the accuracy comparison.

\subsubsection{Effects of embedding size}
The rightmost figures of Fig.~\ref{fig:rmse_aiv} - \ref{fig:n_aiv}
show the recommendation accuracy with the embedding size $d_{\mathrm{user}}$
varying in $\{50,100,200,250,300\}$. In Fig.~\ref{fig:rmse_aiv} (right),
we set $d_\mathrm{dict} = 64$ and $d_\mathrm{GRU} = 32$.
In Fig.~\ref{fig:p_aiv}, we set $d_\mathrm{dict} = 512$
and $d_\mathrm{GRU} = 256$. We can see from the results that higher accuracy
can be achieved with larger user embedding size, which is consistent with
existing methods~\cite{paterek2007improving,koren2008factorization}.
Therefore, we choose $d_{\mathrm{user}} = 300$ in the following accuracy
comparison although higher $d_{\mathrm{user}}$ could further improve the
performance of S2PNM.

\label{sec:qual_anly}
\begin{table}[tb!]
	\caption{Test RMSE comparison between S2PNM and seven state-of-the-art
		rating prediction methods on the Netflix dataset. S2PNM with $d_\mathrm{user}$=50
		can outperform SVD++, TimeSVD++,
		GLOMA and  MRMA of the rank of 300,
		I-AutoRec and NADE with 1000-dimension
		embeddings, RRN with 300-dimension embeddings. We also present the relative performance gain in percentage over SVD++.}
	\centering
	\begin{tabular}{ l | l | c }\hline
		Method&Model&RMSE        \\\hline\hline
		SVD++ &Factorization            &     0.89267 $(+ 0.00\%)$       \\
		TimeSVD++ &Factorization      &     0.90762 $(- 1.67\%)$     \\
		GLOMA &Factorization                        &     0.89326 $(- 0.06\%)$       \\
		MRMA &Factorization                    					   &      0.89210 $(+ 0.06\%)$      \\
		I-AutoRec &Neural                                 &     0.92185  $(- 3.27\%)$     \\
		RRN &Neural                                           &     0.89014  $(+ 0.28\%)$       \\
		NADE(2 layers) &Neural           &	 0.88876  $(+ 0.43\%)$		 \\\hline
		S2PNM, $d_{\mathrm{user}}$=50 &Neural         &    {0.87661} $(+ 1.80\%)$     \\
		S2PNM, $d_{\mathrm{user}}$=200 &Neural       &    {0.87386} $(+ 2.11\%)$     \\
		S2PNM, $d_{\mathrm{user}}$=300 &Neural       &    $\mathbf{0.87322 (+ 2.18\%)}$     \\\hline
	\end{tabular}
	\label{tbl:timeRMSEmp}
\end{table}

\subsection{Rating Prediction Comparison}

\subsubsection{Comparison with State-of-the-art methods}
This experiment compares the rating prediction accuracy of
S2PNM against state-of-the-art factorization
methods~\cite{koren2008factorization,koren2010collaborative,chen2017gloma,mrma}
and neural methods~\cite{autorec,rrn2017,zheng2016neural}.
Note that SVD++~, GLOMA and MRMA are factorization models
which assume that user preferences are static, while TimeSVD++ \cite{koren2010collaborative} used a time-dependent bias term to capture
the temporal effects. RRN~\cite{rrn2017} and NADE~\cite{zheng2016neural}
both leverage the neural autoregressive model to extract the patterns of
how user future actions are affected by his/her historical behaviors.
For S2PNM, we use a single-layer RNN with $256$ GRU units, the dimension of
user embedding $d_{\mathrm{user}}$ ranges in $\{50, 200, 300\}$, and
set the dictionary size as $d_{ \mathrm{dict}} = 1000$.

Table~\ref{tbl:timeRMSEmp} compares the RMSE of all the methods on the
Netflix prize dataset. We can see that S2PNM with $d_{\mathrm{user}}$ = 50
can outperform SVD++\cite{koren2008factorization}, TimeSVD++~\cite{koren2010collaborative},
GLOMA~\cite{chen2017gloma} and  MRMA~\cite{mrma} with the rank of 300,
I-AutoRec~\cite{autorec} and NADE~\cite{zheng2016neural} with 1000-dimension
embeddings, RRN~\cite{rrn2017} with 300-dimension embeddings.
This demonstrates that S2PNM is more effective to predict user ratings than
the compared methods. In addition, compared with NADE, S2PNM achieves $0.43\%$
performance gain and over 5X efficiency improvement. The main reasons why
S2PNM can improve the recommendation accuracy are: 1) the learned dictionary
which maximizes the use of the GRU outputs enriches the expressive capacity
of the S2PNM model, 2) the neural network-based distribution approximator
that attentively reads the dictionary atoms can accurately capture the dynamic
preferences of users, and 3) both the static and dynamic preferences of users
are modelled by S2PNM instead of only modeling the static preferences in
SVD++, GLOMA, MRMA and I-AutoRec.

\subsubsection{Comparison with MRMA on different users}

\begin{figure}[b!]
	\centerline{\includegraphics[width=.45\textwidth]{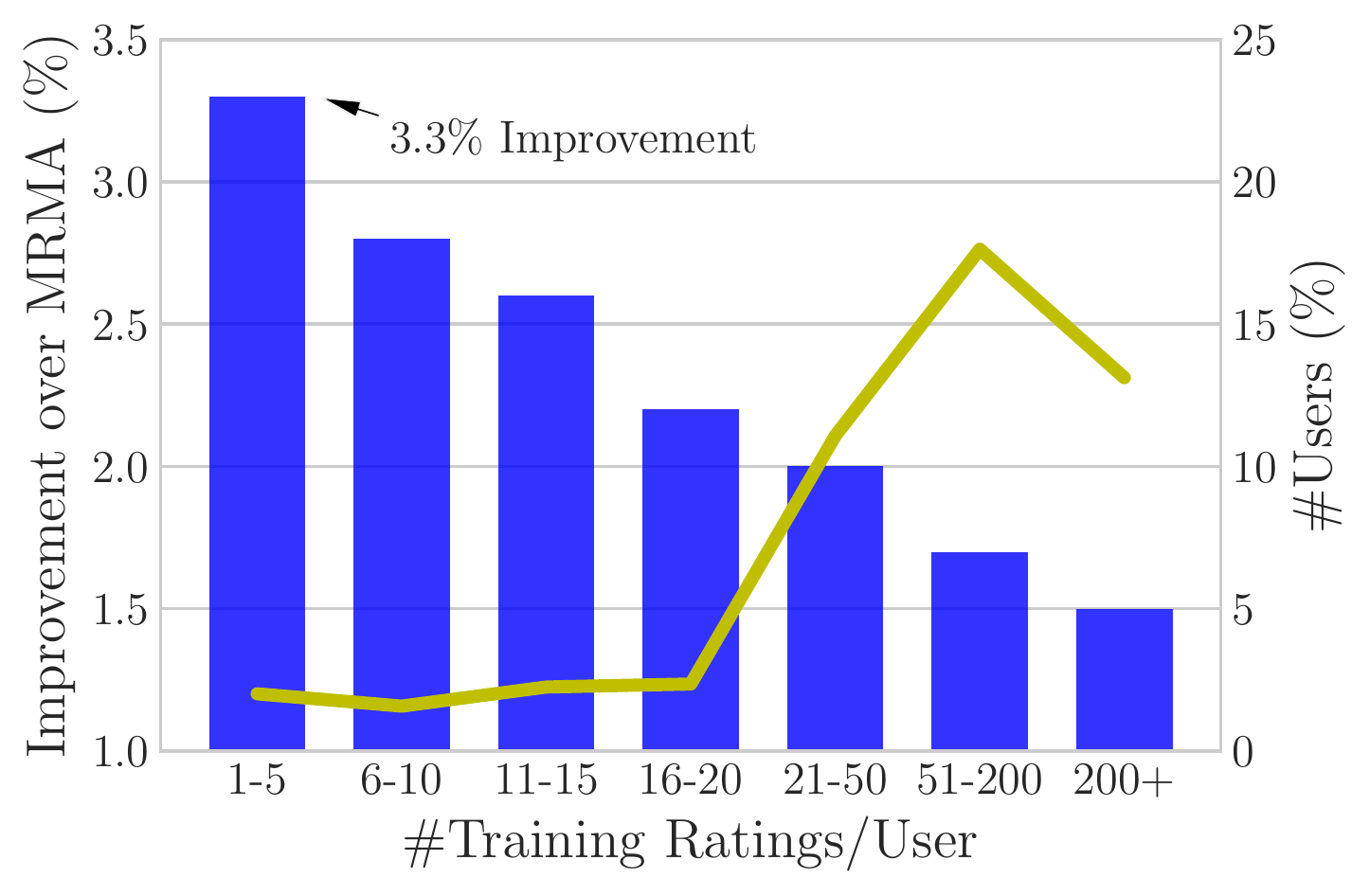}}
	\caption{\label{fig:upgain}
		RMSE improvements over MRMA on users with varying number of training ratings.
		S2PNM achieves $3.3\%$ improvement on users with less than $5$ ratings.
		Users with less ratings benefit more from S2PNM. The yellow line denotes
		the fraction of users.}
\end{figure}

To understand how sequential information help in recommendation,
we conduct a detailed comparison with the MRMA method which only learns
static preferences of users. As shown in Fig.~\ref{fig:upgain}, S2PNM
achieves consistent improvements ($\ge 1.5\%$) over MRMA for all kinds of
users and the largest improvement is $3.3\%$ on users with less than $5$
ratings. The fact that users with less ratings benefit more from S2PNM
indicates that (1) the sequential information captured by S2PNM are
indeed helpful when little information of users are obtained from ratings
(not enough ratings),
and (2) S2PNM may be applied to address the on long-tail
user issue where many existing CF methods may fail~\cite{yin2012challenging}.

\subsection{Item Ranking Comparison}
\subsubsection{Comparison with State-of-the-art methods}
\begin{table}[tb]
	\caption{Item ranking performance on the Amazon Baby Care dataset
		in terms of precision@5, hit-rate@5 (HR@5) and NDCG@5. Higher values
		indicate better performance. Bold face indicates the best performing
		result in each column, and `$\%$' is omitted from all numbers.}
	\centering
	\begin{tabular}{ l | c | c | c }\hline
		Measures(\%) & Precision@5 & HR@5 & NDCG@5         \\
		\hline\hline
		SVD++  & 0.618 & 3.108 & 0.650	     \\
		BPR            &  0.689 & 3.334 & 0.694       \\
		eALS           &  0.677 & 3.262 & 0.725   \\
		NeuMF & 0.802  & 3.579 & 0.779     \\
		RUM  & 0.692  & 3.590 & 0.791	          \\
		SHAN & 0.764  & 3.703 & 0.759 \\
		SASRec & 0.812  & 3.901 & 0.787    \\\hline
		S2PNM, $d_\mathrm{user}$=50       &  0.717  &  3.525  &  0.712    \\
		S2PNM, $d_\mathrm{user}$=200     &  0.812  & {3.960}   &  0.835	 \\
		S2PNM, $d_\mathrm{user}$=300     &  \textbf{0.820} &  \textbf{4.000}   & \textbf{0.835}\\\hline
	\end{tabular}
	\label{tbl:rankingCmp}
\end{table}

\begin{figure}[b!]
	\centerline{\includegraphics[width=.45\textwidth]{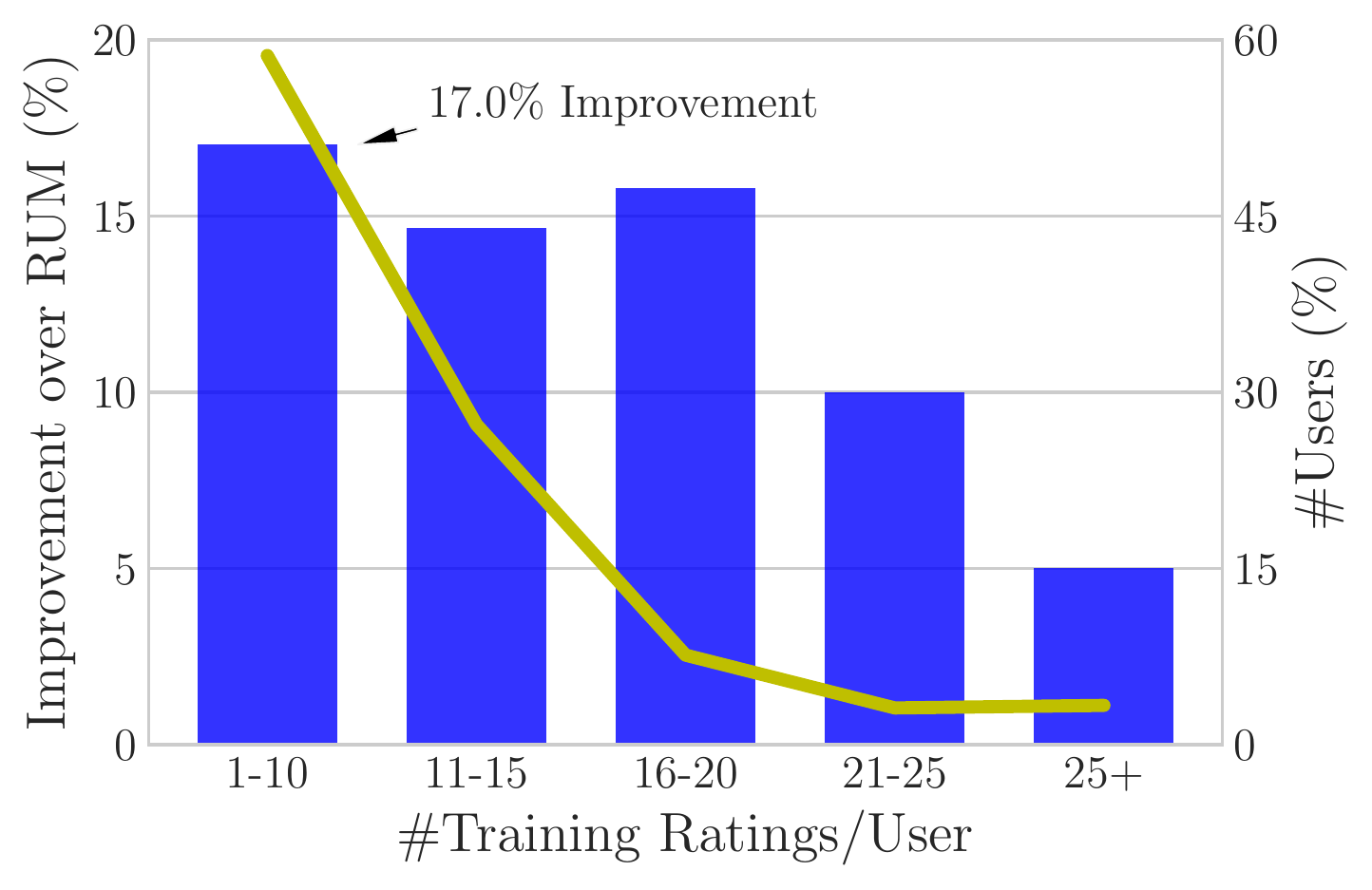}}
	\caption{\label{fig:rgain}
		HR@5 improvements over RUM on users with varying number of training ratings.
		S2PNM achieves $17.0\%$ improvement on users with less than $10$ ratings.
		Users with less ratings benefit more from S2PNM. The yellow line denotes
		the fraction of users.}
\end{figure}

Here, we evaluate S2PNM in
item ranking task. Differing from rating prediction experiment,
this study uses $d_{ \mathrm{dict}}$ = 5000 and $d_{ \mathrm{GRU}}$ = 256
with batch size as 128. Beside SVD++, we compare S2PNM with four ranking-based
methods including BPR~\cite{rendle2009bpr}, eALS~\cite{he2016fast},
NeuMF~\cite{he2017neural}, RUM~\cite{chen2018sequential}, SHAN~\cite{ying2018sequential} and SASREC~\cite{kang2018self}.
Note that NeuMF is a neural method that is designed to utilize implicit feedback
with non-linear interactions to improve performance and RUM is a
state-of-the-art sequential recommendation method based on external memory network
to capture the dynamic user preferences.

Table~\ref{tbl:rankingCmp} shows the recommendation accuracy of S2PNM
and the four compared methods on the Amazon Baby Care dataset. We can
see from the results that the four neural methods (NeuMF, RUM, SHAN and SASREC) significantly
outperform all factorization methods (SVD++, BPR, and eALS), which confirms
that neural methods are indeed more powerful than factorization methods
in sequential recommendation.
As shown in the table, SASREC achieved the best performance among all the compared method.
However, S2PNM outperforms SASREC by 0.99\%, 2.54\% and 6.10\% in terms of Precision\@5, HR\@5 and NDCG@5, respectively. This indicates that S2PNM can better capture the sequential interactions between users and items for more accurate recommendation.

\subsubsection{Comparison with RUM on different users}
As mentioned previously, S2PNM bears some similarities with RUM~\cite{chen2018sequential}. The main difference lies in that RUM directly combines user preferences and sequential patterns, while S2PNM translates sequential patterns to dynamic part of user preferences. Fig.~\ref{fig:rgain} compares S2PNM with RUM by HR@5 (results of NDCG and Precision show the same trend thus omitted for space). It is clear that S2PNM consistently outperforms RUM for all kinds of users and the largest improvement is $ 17.0\%$ on users with less than $10$ ratings. This highlights the importance of translating sequence to preference, especially for sparse users that have few history in training.

\begin{table*}[!t]
	\caption{Understanding the dynamic user preferences captured by S2PNM
		by the changes of neighbors over different periods of time on the Netflix dataset.}
	\centering
	\begin{tabular}{ c | l }
		\hline
		\multicolumn{2}{c}{ neighborhood computed at 1580-\textit{th} day, movies watched during day 1580-1826}	\\\hline
		\tabincell{c}{top three \\nearest neighbors}
		& \tabincell{l}{
			user ${127192}$: star, men, independ, potter, boy, harri, war, sign, harbo\\
			user ${442596}$: star, lord, ring, final, true, king, stor, trek, american, lethal \\
			user ${203167}$: star, man, boy, girl, ring, friend, trek, return, simpson, war}       \\\hline
		\tabincell{c}{top seven movies\\watched by user 1}
		& \tabincell{l}{
			1. Star Wars: Episode VI: Return of the Jed\\
			2. Lord of the Rings: The Two Tower\\
			3. Lord of the Rings: The Fellowship of the Ring\\
			4. Star Wars: Episode IV: A New Hop\\
			5. Harry Potter and the Sorcerer's Stone\\
			6. Pulp Fiction\\
			7. The Godfather
		}\\\hline
		\hline
		\multicolumn{2}{c}{ neighborhood computed at 1826-\textit{th} day, movies watched during day 1826-1860}                 \\\hline
		\tabincell{c}{top three \\nearest neighbors}
		& \tabincell{l}{
			user ${388153}$: citi, sex, friend, star, ring, war, best, lord, return, man\\
			user ${461900}$: soprano, seinfeld, halen, shield, simpson, godfath, pretti\\
			user ${203167}$: star, man, boy, girl, ring, trek, return, last, war, king}       \\\hline
		\tabincell{c}{top seven movies\\watched by user 1}
		& \tabincell{l}{
			1. The Bourne Supremac \\
			2. Harry Potter and the Prisoner of Azkaba\\
			3. Along Came Poll\\
			4. Absence of Malic\\
			5. Road to Perditio\\
			6. 50 First Date\\
			7. Bad Boys\\
		}\\\hline
		\hline
		\multicolumn{2}{c}{ neighborhood computed at 1997-\textit{th} day, movies watched during day 1997-2118}	\\\hline
		\tabincell{c}{top three \\nearest neighbors}
		& \tabincell{l}{
			user ${229573}$: star, ring, trek, lord, black, war, stargat, back, stor, men\\
			user ${203167}$: star, man, boy, girl, ring, friend, trek, return, simpson, war\\
			user ${201895}$: soprano, war, star, bad, sex, godfath, man, citi, die}   \\\hline
		\tabincell{c}{top seven movies\\watched by user 1}
		& \tabincell{l}{
			1. Star Wars: Episode I: The Phantom Menac\\
			2. Lord of the Rings: The Fellowship of the Ring: Extended Editio\\
			3. Lord of the Rings: The Two Towers: Extended Editio\\
			4. Lord of the Rings: The Return of the King: Extended Editio\\
			5. Harry Potter and the Chamber of Secret\\
			6. The Sopranos: Season (itemid: 8116, 5760, 11662, 14302 )\\
			7. Friends: Season (itemid: 2942, 7158, 9909)
		}\\\hline
	\end{tabular}
	\label{tbl:qualEmp}
\end{table*}

\subsection{Qualitative Analysis}
\label{sec:quat_anly}
Here, a qualitative analysis is conducted to better
understand the dynamic user preferences captured by S2PNM.
To verify if the additive user dynamic preference $\overline{u}_t$
captured by S2PNM can help track user preference
drifts and predict user future preferences,
we select the most similar neighbors using Cosine similarity given each
$\overline{u}_t$. Then, we study the semantic meanings of the neighbors
by extracting the most frequent keywords from the movie titles watched by
the neighbors.

Table~\ref{tbl:qualEmp} presents an example for the Netflix user with id=1,
and meanwhile presents the user preference drifts represented by the
changes of neighbors over different periods of time. We can see from the results:
\begin{enumerate}
	\item before day $1580$\footnote{The elapsed day is defined as the number
		of days after the system's initial date -- Nov. 11th 1999.}, we can see that
	the user ids of the top 3 neighbors are: ${127192}$, ${442596}$, ${203167}$
	-- all with strong interests in \emph{Lord of Rings} and \emph{Star War}.
	This suggests that user 1 prefered sci-fiction and fantasy movies,
	which can be verified by the most popular movies watched by user $1$ within
	the period (day 1580 -- 1826);
	
	\item during day 1826 -- 1860, user preferences shifted from sci-fiction and
	fantasy movies to romance and comedy movies, e.g., \emph{Absence of Malic},
	and crime movies, e.g., \emph{Along Came Poll}.
	Similar patterns also occur in neighborhood:
	user ${1}$ was closer to user ${388153}$ who favored romance and comedy movies
	and user ${461900}$ who favored crime movies;
	
	\item at day 1997, the neighbors became comprehensive -
	user ${229573}$ and user ${203167}$ are fans of sci-fiction and fantasy movies,
	user ${203167}$ also shows interests to romance movies, and user ${201859}$ is
	interested to crime movies. Conformed with the patterns unveiled by the neighbors,
	user ${1}$ watched movies across multiple genres in the next 4 months,
	including fantasy movies - \emph{Lord of the Rings} and \emph{Harry Potter},
	sci-fiction movie - \emph{Star War}, romance movie - \emph{Friends}, and
	crime movie - \emph{The Sopranos}.
\end{enumerate}

It is worth noting that traditional matrix factorization methods
work with static user embeddings and thus cannot adjust
user preferences over time. This study confirms that S2PNM can adapt
automatically to the changes of user interest and thus can help adjust
the predictions. We can also learn from the study that user preference
drifts are complex and using simple temporal bias terms, e.g.,
TimeSVD++~\cite{koren2010collaborative}, could not optimally capture
these dynamic information.

\section{Related Work}
\label{sec:related}
	
Many collaborative filtering works~\cite{Adomavicius05,Su09} formulate personalized recommendation problems as matrix completion problems, of which the goal is to recover the missing entries in the rating matrix based on low-rank assumptions. In general, the attributes or preferences of a user are modeled by linearly combining item factor vectors using user-specific coefficients. And most of traditional CF solutions \cite{rennie2005fast,singh2008relational,koren2009matrix,yu2009fast,rendle2009bpr,lee2013local,Chen15,he2016fast} assume the user profiles and item attributes are \textit{static} so that temporal or sequential information are ignored. Probably the most popular variants are Probabilistic Matrix Factorization (PMF)~\cite{mnih2007probabilistic} and its Bayesian extension~\cite{salakhutdinov2008bayesian}, which achieved robust and strong results in rating prediction. In addition to simple matrix factorization based CF models, hybrid methods have also been investigated in the literature. The Netflix Prize winners Bell~\etal~\cite{bell2007modeling} and Koren~\etal~\cite{koren2008factorization} utilized the combination of memory-based and matrix factorization methods to improve the recommendation accuracy. Another research line focuses on the issue of the computational efficiency, for example Mackey~\etal~\cite{mackey2011divide} employed a Divide-Factor-Combine (DFC) framework as well as~\cite{lee2013local,Chen15,Wang16BPMTMF}, in which the expensive task of matrix factorization is randomly divided into smaller subproblems which can be solved in parallel using arbitrary matrix factorization algorithms.

Temporal aspects in recommendation were discussed in TimeSVD++~\cite{koren2010collaborative}. The key innovation here lies in that TimeSVD++ introduced time-dependent bias terms to capture temporal dynamics caused by rating scale changes and popularity changes in an integrated fashion~\cite{rrn2017}. However, the features of TimeSVD++ are hand engineered similar to SVDFeature~\cite{chen2012svdfeature}, which makes the model difficult to adapt to new problems due to the lack of the knowledge of the new dataset. To remedy this issue, Factorized Personalized Markov Chains (FPMC)~\cite{RendleWWW10} and its extension -- hierarchical representation model (HRM)~\cite{WangSIGIR15} were proposed, of which both embedded the adjacent behavior transition into the latent space. By doing so, the local patterns between items can be exploited in an end-to-end way. Beyond two-step behavior modeling, Markov chain is used for modeling sparse sequence\cite{HeICDM16}. One major issue of using Markov chains is potential state space explosion in face of different possible sequences over items.

As the revolution of deep learning, many efforts~\cite{guo2017deepfm,he2017neural,lian2018xdeepfm} have been made to adopt neural networks to solve the recommendation tasks, which are also mostly focused on the static collaborative filtering setting, i.e., without considering the temporal or sequential information. In specific, one of the earliest works proposed to apply Restricted Boltzmann Machines (RBM)~\cite{SalakhutdinovICML07} for collaborative filtering. Later, Autoencoder-based method~\cite{autorec} was proposed, which regards the recommendation task as a denoising problem. To exploit the ordinal nature of ratings, neural auto-regressive distribution estimator (NADE) \cite{zheng2016neural} has been used to perform collaborative filtering. The above neural methods showed excellent empirical performance on popular 	benchmark datasets in static recommendation.
	
More recently, another line of emerging works employ recurrent/memory networks~\cite{hochreiter1997long,cho2014learning,ma2019hierarchical} to tackle the more practical task of sequential recommendation -- to predict the future behavior given a user's historical rating records. This family of the algorithms \cite{YuSIGIR16,chen2018sequential,huang2018isrkemn,rrn2017,ying2018sequential} can usually outperform the aforementioned Markov chain-based methods~\cite{RendleWWW10,WangSIGIR15} due to the higher model capacity. In more detail, the Dynamic REcurrent bAsket Model (DREAM)~\cite{YuSIGIR16} embeded user historical ratings by a RNN to predict his/her future preference. And similarly the Recurrent Recommender Network (RRN)~\cite{rrn2017} proposed to employ LSTM~\cite{hochreiter1997long} to capture the dynamics of both users and items. Complementary to the RNN-based methods, Huang~\etal~\cite{huang2018isrkemn} and Ren~\etal~\cite{ren2019lifelong} alternatively adopted memory networks~\cite{graves2014neural,sukhbaatar2015end} to learn the short-term patterns, together with item attributes. In the meanwhile, the Sequential Hierarchical Attention Network (SHAN)~\cite{ying2018sequential} and \cite{kang2018self,zhang2019feature} introduced the attention mechanism to automatically assign different influences of items in a user's long-term set so that the dynamic properties can be captured, then relied on another attention layer to couple user sequential behavior with long-term representation.

Probably most closely related to our work is RUM~\cite{chen2018sequential}, one of the few neural networks models for sequential recommendation. It utilized an external memory matrix to maintain item-level historical information. When making predictions, the memory of the latest interacted items in a fixed-size window (controlled by parameter $K$) would be attentively read out to generate an embedding as dynamic part of the user representation. However, we argue that the different sequential patterns might emerge on different timescales such that a simple first-in-first-out windowed mechanism might limit the model capacity. From the experiments, we can observe that our model outperforms state-of-the-art RUM in term of ranking prediction.

In a nutshell, this work formulates the sequential recommendation task as a supervised dictionary learning task. The proposed S2PNM method learns a dictionary to construct user dynamic preferences, which can embed user static and dynamic preferences under the same latent space, making the model very compact. Thus, S2PNM can achieve superior performance by a simple additive mechanism to fuse the static and dynamic preferences of users. To the best of our knowledge, this is the first work that translates user rating sequence into user preference via dictionary learning for sequential recommendation. The experiments on Netflix and Amazon datasets demonstrate that S2PNM can significantly outperform the state-of-the-art matrix factorization methods and neural collaborative filtering methods in the realistic setting.

\section{Conclusion and Future Work}
\label{sec:conclsn}
This paper proposes S2PNM -- a sequence-to-preference neural machine for sequential recommendation. In particular, we propose the Seq2Pref Network for dynamic user preference modeling, which first embeds the sequential dependencies into a latent vector and then translates the dynamic preference into the latent space of user static preference using a learned dictionary. Empirical studies on multiple real-world datasets demonstrate that S2PNM can achieve significantly higher accuracy compared with state-of-the-art factorization and neural sequential recommendation methods.

One line of future work is to study pair-wise learners for S2PNM and extend S2PNM for multi-media items, whereby multi-media items such as videos and images, contains much richer semantics that reflects the interest of users. This requires the model capable of modeling auxiliary information -- learn from multi-view and multi-modal data. The second future work is to introduce non-linearity when combing the static user preferences and dynamic user preferences, which may further improve the performance of the proposed method~\cite{He17Sparse}. Another emerging problem is to learn the meaningful dependencies from the fragmented sequence instead of overall sequences, without sacrificing the recommendation qualities. This matters a lot for practical recommender systems. One more future direction is to explore the potential of time-aware recurrent neural networks by devising an extra loss of the RNN-based auto-regressive models to improve the performance of event-time prediction.
	
\section*{Acknowledgements}
This work was partially supported by National Natural Science Foundation of China under Grant Nos. U19B2035 and 61972250, and National Key Research and Development Program of China under Grant Nos. 2016YFB1001003, 2018AAA0100704 and 2018YFC0830400.

\bibliographystyle{abbrv}
\bibliography{main}

\begin{IEEEbiography}
	[{\includegraphics[width=1in,height=1.25in,clip,keepaspectratio]{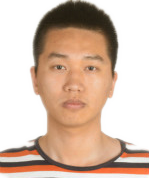}}]
	{Chao Chen} (M'19) is a PhD candidate in School of Electronic Information and Electrical Engineering, Shanghai Jiao Tong University, Shanghai, China. He joined IBM Research - China since May 2016, and before that he studied in department of Computer Science from Tongji University, Shanghai, China. His research interests focus on applying neural networks techniques to recommender systems.
\end{IEEEbiography}

\begin{IEEEbiography}
	[{\includegraphics[width=1in,height=1.25in,clip,keepaspectratio]{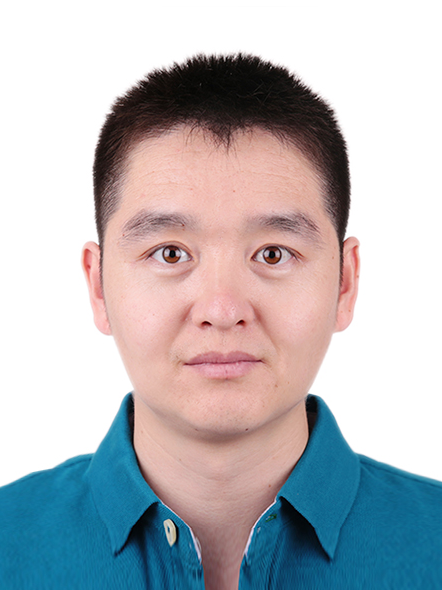}}]
	{Dongsheng Li} (M'14) is a senior researcher with Microsoft Research Asia (MSRA), Shanghai, since February 2020.  Before joining MSRA, he was a research staff member with IBM Research -- China since April 2015. He is also an adjunct professor with  School of Computer Science, Fudan University, Shanghai, China. He obtained Ph.D. from School of Computer Science of Fudan University, China, in 2012. His research interests include recommender systems and  general machine learning applications.
\end{IEEEbiography}

\begin{IEEEbiography}[{\includegraphics[width=1in,height=1.25in,clip,keepaspectratio]{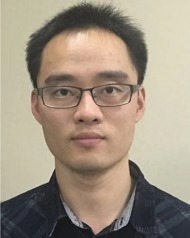}}]{Junchi Yan} (M'10) is currently an Associate Professor (PhD Advisor) with Department of Computer Science and Engineering, and AI Institute of Shanghai Jiao Tong University. He is also the co-director for the prestigious SJTU ACM Class (in charge of AI direction). Before that, he was a Senior Research Staff Member with IBM Research -- China where he started his career since April 2011, and once an adjunct professor with the School of Data Science, Fudan University. His research interests are machine learning and computer vision. He serves as Associate Editor for IEEE ACCESS, (Managing) Guest Editor for IEEE Transactions on Neural Network and Learning Systems, Pattern Recognition Letters, Pattern Recognition, Vice Secretary of China CSIG-BVD Technical Committee, and on the executive board of ACM China Multimedia Chapter. He has published 40+ peer reviewed papers in top venues in AI and has filed 20+ US patents. He has once been with IBM Watson Research Center, Japan NII, and Tencent/JD AI lab as a visiting researcher. He won the Distinguished Young Scientist of Scientific Chinese and CCF Outstanding Doctoral Thesis.
\end{IEEEbiography}

\begin{IEEEbiography}[{\includegraphics[width=1in,height=1.25in,clip,keepaspectratio]{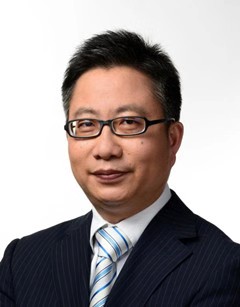}}]{Xiaokang Yang} (M'00-SM'04-F'19) received the B. S. degree from Xiamen University, in 1994, the M. S. degree from Chinese Academy of Sciences in 1997, and the Ph.D. degree from Shanghai Jiao Tong University in 2000. He is currently a Distinguished Professor of School of Electronic Information and Electrical Engineering, Shanghai Jiao Tong University, Shanghai, China. His research interests include visual signal processing and communication, media analysis and retrieval, and pattern recognition. He serves as an Associate Editor of IEEE Transactions on Multimedia and an Associate Editor of IEEE Signal Processing Letters. Prof. Yang is also a fellow of IEEE.
\end{IEEEbiography}

\end{document}